\documentclass[a4paper,fleqn,usenatbib]{mnras}
\usepackage{graphicx}
\usepackage{subfigure}
\usepackage{amsmath}
\usepackage{amssymb}
\usepackage{multirow}
\usepackage[mathscr]{euscript}
\usepackage[para,online,flushleft]{threeparttable}
\usepackage{multirow}
\usepackage{booktabs}
\usepackage{xr}
\usepackage{array}
\usepackage{graphicx}
\usepackage[dvipsnames]{xcolor}
\usepackage{multicol}
\usepackage{colortbl}
\usepackage{xcolor}

\newcolumntype{L}[1]{>{\raggedleft\arraybackslash}m{#1}}

\newenvironment{rcases}
  {\left.\begin{aligned}}
  {\end{aligned}\right\rbrace}

\title[Radio spectra of protostellar jets]{Radio spectra of protostellar jets: Thermal and non-thermal emission}

\author[Mohan et al.]{Sreelekshmi Mohan\thanks{E-mail: sreelekshmimohan@res.iist.ac.in}, S. Vig and S. Mandal\\
Indian Institute of Space Science and Technology, Thiruvananthapuram, 695 547, India\\
}

\date{Accepted XXX. Received YYY; in original form ZZZ}

\pubyear{}
\begin{document}

\date{}

\pagerange{\pageref{firstpage}--\pageref{lastpage}} \pubyear{}

\maketitle

\label{firstpage}

\begin{abstract}
Protostellar jets and outflows are pointers of star-formation and serve as important sources of momentum and energy transfer to the interstellar medium. Radio emission from ionized jets have been detected towards a number of protostellar objects. In few cases, negative spectral indices and polarized emission have also been observed suggesting the presence of synchrotron emission from relativistic electrons. In this work, we develop a numerical model that incorporates both thermal free-free and non-thermal synchrotron emission mechanisms in the jet geometry. The flux densities include contribution from an inner thermal jet, and a combination of emission from thermal and non-thermal distributions along the edges and extremities, where the jet interacts with the interstellar medium. We also include the effect of varying ionization fraction laterally across the jet. An investigation of radio emission and spectra along the jet shows the dependence of the emission process and optical depth along the line of sight. We explore the effect of various parameters on the turnover frequencies and the radio spectral indices (between 10~MHz and 300~GHz) associated with them.
\end{abstract}


\section{Introduction}

Jets are the consequence of accretion of material in a rotating system. From ultra-relativistic/relativistic jets in high energy astrophysical phenomena \citep{doi:10.1146/annurev.astro.37.1.409,doi:10.1146/annurev-astro-081817-051948} to non-relativistic jets in protostars \citep{Lee_2007, Reipurth_2004} and brown dwarfs \citep{2009ASSP...13..259W,10.1093/mnras/stu1461}, jets assist in accretion by carrying away excess angular momentum \citep{1982MNRAS.199..883B,1983ApJ...274..677P}. Jets serve as an important tool in the study of early stages of stellar birth, because newly forming stars are embedded in large clouds of gas and dust making them observationally challenging to probe. However, jets and the gas that they entrain in the form of outflows are easier to detect because of their kinematic signatures. Protostellar jets represent a significant signpost of ongoing star formation \citep{2006A&A...453..911F} as they influence the surrounding interstellar medium (ISM) by the transfer of momentum and energy. Protostellar jets can either augment or retard the star-formation occurring in their vicinity \citep{2006ApJ...640L.187L, 2007ApJ...662..395N, 2008ApJ...683..255S}. 

\par Although the jet launching mechanism is still debated, jets are believed to be closely linked to the magnetic field structure of the associated protostellar system. The collimation of jets is achieved by the confining pressure of toroidal component of helical magnetic fields \citep{1997ASPC..121..845L,Meier84} that has been established by various hydromagnetic and magnetohydrodynamic simulations \citep{1987ApJ...315..504L,1995ApJ...439L..39U, Cerqueira_2001,zanni2004mhd,Bellan2005}. Observationally, the presence of magnetic field in protostellar jets have been investigated using methods such as spectro-polarimeteric observations \citep{2010MNRAS.409.1347D}, and mapping of spectral line polarization \citep{2018NatCo...9.4636L}, to name a few. 
The presence of synchrotron emission can also be used to gauge the direction and strength of magnetic field. It is interesting to note that theories predict similar morphologies for jet magnetic fields, irrespective of whether they are of protostellar or AGN origin  \citep{1986NYASA.470...88K,2000AIPC..522..275L}. 
Sensitive observations have recently revealed not only the presence of polarized emission confirming synchrotron emission towards a massive protostellar jet \citep{2010Sci...330.1209C}, but also the co-location of synchrotron emission along the edges and termination points of the jet, where the interaction of strong shocks against the ambient medium is likely to produce efficient particle acceleration \citep{2017ApJ...851...16R}. 

\par The theory of synchrotron emission from high energy phenomena related to extra-galactic jets is well-established \citep{1982ApJ...256...13R,1987ApJ...322..643B,1998ApJ...497L..17S} because of the large volume of observational evidence that is available \citep{2006ARA&A..44..463H,2000ApJ...543..373D,1999ApJ...523..177W}. However, there is lack of clarity regarding the origin of synchrotron emission from protostellar jets. This is mainly owing to the fact that there are fewer detections of synchrotron emission from protostellar jets, which in turn is related to their lower energies. 
The best studied case is that of protostellar jet HH~80-81 from a massive protostar which is one of the most collimated jets in our Galaxy \citep{2010Sci...330.1209C}. High frequency observations of this jet have revealed negative spectral indices in the jet and associated knots \citep{1989RMxAA..17...59R,1993ApJ...415..191C,1993ApJ...416..208M}, and low frequency radio observations made it easier to detect non-thermal emission \citep{2018MNRAS.474.3808V}. Among the lower mass counterparts, DG-Tau stands out in terms of non-thermal emission from protostellar jet and observations across a range of radio frequencies \citep{2014ApJ...792L..18A,2018MNRAS.481.5532P,2019ApJ...885L...7F}.

\par A review of literature shows that the radio spectral indices from protostellar jets are usually modeled as thermal free-free emission or synchrotron emission or a combination of both using simplistic considerations of uniform electron density, temperature, etc \citep{2011IAUS..275..367R,2016MNRAS.460.1039P,2018A&ARv..26....3A}. We are interested in constructing a model of the radio jet that can encompass both the thermal as well as non-thermal emission as a function of various parameters such as the opening angle, density and ionization fraction across the jet. For thermal jets, the most widely employed model is that of \citet{1986ApJ...304..713R}, hereafter Reynolds model, which analytically calculates the radio emission and spectral indices using thermal free-free emission for various jet geometries. 
The model developed in the current work involves a more general geometry with additional improvements, described in the next section. The motivation is to comprehend the nature of emission mechanism from a protostellar jet through radio spectral indices in a more realistic scenario than what is generally assumed. 

\par The organization of the paper is as follows. In Sect.~\ref{model}, we introduce our model and describe the model geometry. This is followed by a discussion of the radio emission mechanisms considered in the model and the associated model spectra in Sect.~\ref{sec:emission}. The lateral variation of the ionization fraction introduced in our model is presented in Sect.~\ref{ionizefrac}. The effects of various model parameters on the radio spectra are investigated in Sect.~\ref{ap:spec_free}. Finally, we present a short summary in Sect.~\ref{summary}. 

\section{The Model} \label{model}

\par In order to simulate the  spectrum of a protostellar jet, we formulate a model to explain the observed radio spectral indices using thermal and non-thermal emission mechanisms. The Reynolds model incorporated free-free emission from a highly collimated ionized thermal jet i.e. small opening angles are assumed. In addition, the flux densities are calculated under the assumption that the jet is optically thick ($\tau$ $>>1$) towards the base and optically thin ($\tau$ $<<1 $) at farther radial distances. In other words, the Reynolds model assumes that the total flux density at a given frequency from the full length of the jet is the sum of fully optically thin and fully optically thick flux densities at that frequency. The significant improvements of our model over the Reynolds model, include the following.
\begin{enumerate}
 \item We incorporate a more general geometry and do not assume small opening angle for the jet.
  \item A significant addition is the inclusion of non-thermal synchrotron emission. In order to incorporate this, we take insights from observations of protostellar jets where the synchrotron emission is observed from the edges and termination regions of the jet \citep{{2017ApJ...851...16R},{2016ApJ...818...27R}}.
 \item We have also introduced the lateral variation of ionization fraction across the jet cross-section. At any given length, the jet has maximum ionization fraction along the long central-axis and the fraction of ionized gas decreases with distance from the axis outwards to the cross-sectional edge.
 \item We have accommodated intermediate optical depth values corresponding to intermediate radial distances.
 
\end{enumerate}

Our model has a two-fold application. This can be employed to model radio emission from (i) thermal jets at smaller distances (i.e $r_0 \simeq$ $10$~au), as well as (ii) knots observed farther away from Young Stellar Objects (YSOs).  
We bring to attention the fact that our model does not include the effects of dust emission and for higher radio frequencies ($\gtrsim50$~GHz), this contribution may need to be included.

\begin{figure}
	\includegraphics[width=1.1\columnwidth]{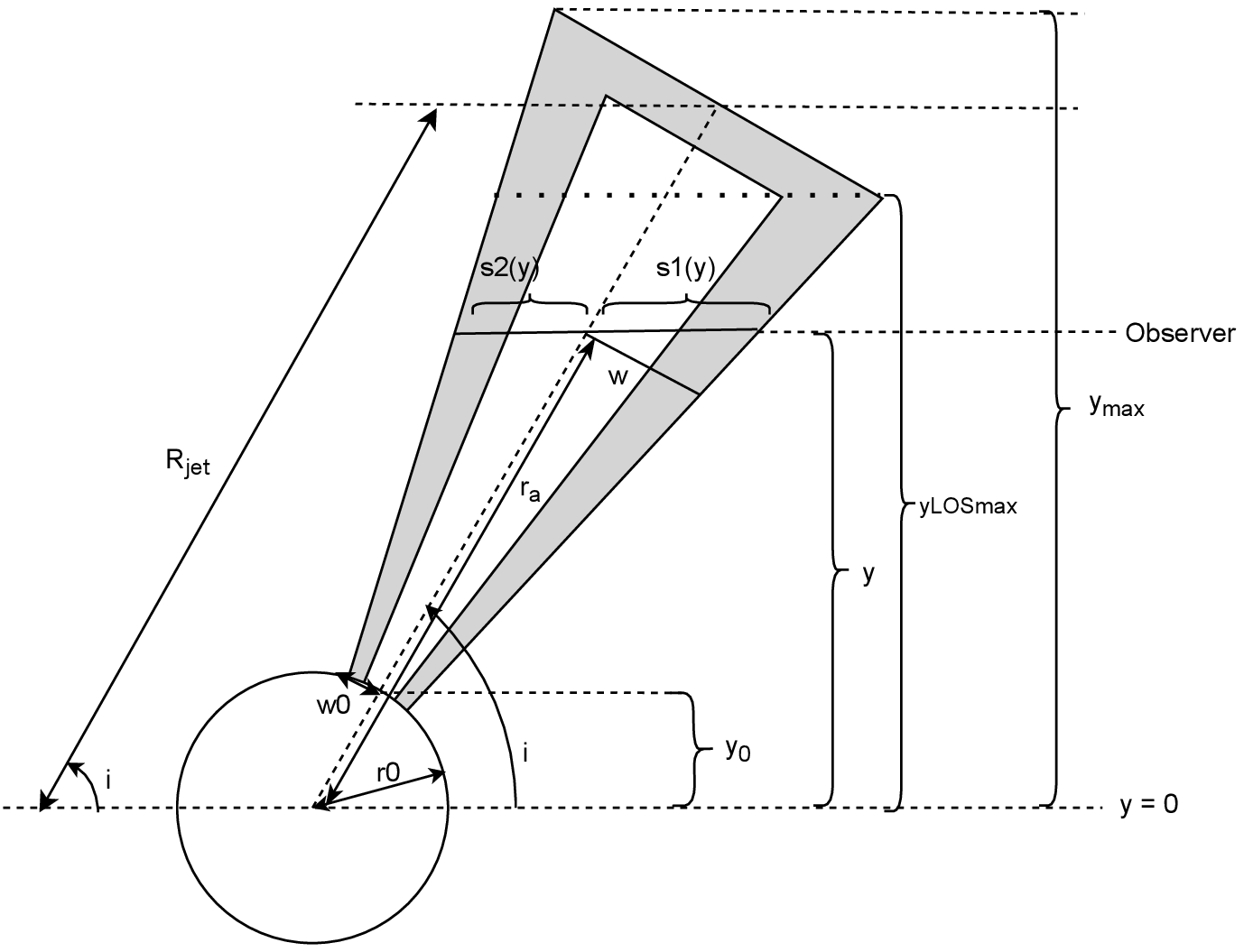}
    \caption{Schematic diagram of a jet with constant opening angle, following the schematic of \citet{1986ApJ...304..713R}. The unshaded jet region represents the inner (near the jet-axis) fully thermal jet. The shaded area represents the geometrically thin region with shocked material that contributes to a combination of thermal and non-thermal emission. The projected distance corresponding to the jet injection radius is marked as $y_0$.}
    \label{fig:geom}
\end{figure}

\subsection{Jet description and geometry} \label{desc_geom}

\par The geometry and orientation of the jet with respect to the observer is shown in Fig.~\ref{fig:geom}. We assume the injection of jet at radius $r_0$, with an opening angle $\theta_{0}$, and having a half-width $w_{0}$ at $r_0$. $r_0$ represents the distance from the central source where the jet is sufficiently ionized for free-free emission to be detectable. In addition, a thin outer shell (shaded region in Fig.~\ref{fig:geom}) of material that emits a combination of thermal free-free and non-thermal synchrotron emission is incorporated. In our calculations we assume that for any point considered within the jet, the distance from that point to the central star is given by $r$ and the points with the same $r$ are located on a spherical surface, centered on the star. For each of the points along a given LOS, the radial distances from the star are different. The maximum length of the jet as measured along the axis is $\text{R}_{\text{jet}}$. The jet is inclined at an angle $i$ with respect to the line-of-sight (LOS) direction of the observer. The jet model shown in the figure and discussed throughout this paper considers a single lobe of a bipolar jet that is blue-shifted. The analysis and results are similar for the red-lobe. 

\par We define the radial distance measured along the axis of the jet as $r_{a}$. The length of the jet projected in the plane of the sky, $y$, and the half-width of the jet ($w$) at any radial distance $r_{a}$, are given by the following relations.

\begin{equation}\label{eq:y}
y = r_{a}\, \sin{i}
\end{equation}
\begin{equation}\label{eq:width}
w(y) = w_{0} \bigg(\frac{r_{a}}{r_{0}}\bigg)^{\epsilon} = w_{0} \bigg(\frac{y}{y_{0}}\bigg)^{\epsilon}
\end{equation}

\par For a given $y$, in addition to $w$, another important parameter is the LOS jet cross-section. In Fig.~\ref{fig:geom}, $\mathscr{S} = s_{1}(y) + s_{2}(y)$ represents the LOS cross-section across the center of a jet. As evident from the figure, $s_{1}(y)$ and $s_2(y)$ represent the widths of the jet on the front and rear side of the jet long axis (with respect to the observer), respectively. The dotted line in the figure shows the maximum LOS distance to the observer through the jet material, perpendicular to the sky plane. This means that the column of jet material contributing to the emission is the maximum at the projected distance $y_{\text{LOS}_{\text{max}}}$. The length $y_{max}$ shown in the figure corresponds to the farthest tip of the projected jet, and is given by the following equation.
\begin{equation}
y_{max} = \text{Y}_{\text{jet}} + w(\text{Y}_{\text{jet}})\cos{i}
\end{equation}

\noindent where, $\text{Y}_{\text{jet}}$ = $\text{R}_{\text{jet}} \sin{i}$. The jet geometry is dictated by the nature of evolution of the jet width $w$ with the jet-axis radial distance $r_{a}$ from the injection distance $r_0$. This is determined by the power-law index $\epsilon$ defined in Eqn.~\eqref{eq:width}. For instance, $\epsilon=1$ represents the case of a conical jet with a constant opening angle, and $\epsilon=0$ would imply a constant width throughout the jet, suggesting a highly collimated nature of the jet. Values with $\epsilon > 1$ would lead to increase in opening angle with radial distance.

To understand the effect of $\epsilon$ on the jet collimation, we consider the variation of jet opening angle ($\theta$) as a function of jet-axis radial distance $r_{a}$ for different values of $\epsilon$. This is given by the following expression.
\begin{equation}\label{eq:op_ang}
\tan{(\theta/2)} = \frac{w}{r_{a}} = \frac{w_o}{r_{a}}\left(\frac{r_{a}}{r_0} \right)^\epsilon = \tan{(\theta_{0}/2)}\left(\frac{y}{y_{0}}\right)^{\epsilon-1}
\end{equation}

\begin{figure}
	\hskip -0.4cm
	\includegraphics[width=\columnwidth]{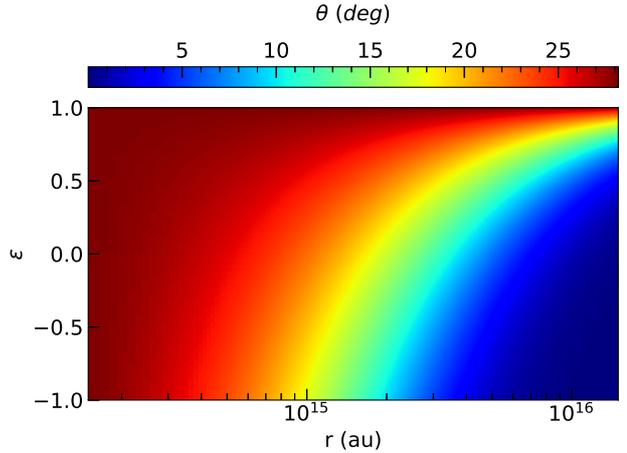}
	\caption{Jet opening angle, $\theta$, as a function of power-law index of jet width ($\epsilon$) and jet-axis radial distance ($r_{a}$). }
    \label{fig:epsilon_jetcoll}
\end{figure}

We consider a protostellar jet with $r_{0} = 10$~au, which is reasonable since high resolution observations have probed jets very close to the exciting source \citep{{2017NatAs...1E.152L},{2019A&A...631A..64B},1998Natur.396..650G,2009A&A...494..147L,2010ApJ...708...80M}. Further, we assume $\text{R}_{\text{jet}} = 1000$~au measured along the axis and inclination $i=90^\circ$. The jet collimation as a function of $\epsilon$ along the jet-axis radial distance $r_{a}$ is displayed in Fig.~\ref{fig:epsilon_jetcoll}. It can be seen that negative values of $\epsilon$ imply the contraction of jet width at outer radial distances i.e. to higher values of $y$. The conical case of constant opening angle for $\epsilon=1$ can be clearly visualized.


\section{Emission from the model jet} \label{sec:emission}

\subsection{Free-Free emission } \label{ff}

\par We first present a description of the calculations involved in the free-free emission model. We have already introduced the initial jet injection radius $r_{0}$, initial jet half-width $w_{0}$ and the opening angle $\theta_{0}$. The other physical quantities such as jet velocity ($v$), temperature ($T$), electron number density in the jet material ($n$) and ionization fraction ($x$) have radial power-law indices given by $q_{v}$, $q_{T}$, $q_{n}$ and $q_{x}$, respectively. Note that $q_v$ is related to $q_n$ by the mass conservation equation.
\begin{equation}\label{eq:q_v}
q_v +q_n+2\epsilon=0
\end{equation}
The free-free emission and absorption coefficients are given by the following expressions.
\begin{equation}\label{ff_jnu}
j^{ff}_\nu\, = a_j\, n^2\, x^2\, T^{-0.35}\, \nu^{-0.1}
\end{equation}
\begin{equation}\label{ff_alphanu}
\alpha^{ff}_\nu\,=a_k\, n^2\, x^2\, T^{-1.35}\, \nu^{-2.1}
\end{equation}
\noindent Here, $a_j\,=6.50 \times 10^{-38}\,$~ergs~cm$^{3}$~Hz$^{-0.9}$~K$^{0.35}$~s$^{-1}$~sr$^{-1}$ and $a_k\,=0.212$~cm$^{5}$~K$^{1.35}$~Hz$^{2.1}$ are the proportionality constants of emission and absorption coefficients, respectively, in cgs units. From Eqns.~\eqref{ff_jnu} and \eqref{ff_alphanu}, it is evident that $j^{ff}_{\nu}$ and $\alpha^{ff}_{\nu}$ are functions of number density and temperature. Therefore, a radially varying density or temperature profile results in a variation of emission and absorption coefficients along a LOS. Assuming a LOS variable $s$, the source function of the jet material is given by $\frac{j^{ff}_{\nu}(s)}{\alpha^{ff}_{\nu}(s)}$. The flux density from a LOS of the jet that subtends an incremental solid angle d$\Omega$ at the observer's location, is given by the following expression.

\begin{equation}\label{eq:fluxGeneral}
S_{\nu} = \int_{y_{0}}^{y_{max}} d\Omega\, \int_{0}^{\tau_{\nu}} \frac{j^{ff}_{\nu}(s)}{\alpha^{ff}_{\nu}(s)}\, (e^{-(\tau^{ff}_{\nu} - \tau (s)) })\, d\tau
\end{equation}
Here, $\tau_{\nu}^{ff}\, = \, \int \alpha_{\nu}(s) \, ds$ and $\tau(s)$ are the total LOS optical depth and the optical depth corresponding to any $s$ along a LOS, respectively. The second integral in Eqn.~\eqref{eq:fluxGeneral} incorporates the radial dependence of physical parameters in the volume element along a LOS. This model accommodates a large opening angle in contrast with the Reynolds model which deals with narrow jets and to allow for this, we integrate at each location along the width of the jet (on the plane of the sky). This implies an integration along the LOS of $\mathscr{S}$ which is maximum along the jet axis to zero at the edges.

The integration over $y$ is carried out to estimate the flux density from the entire length of the jet.
\begin{equation}\label{eq:fluxdbl_int}
S_{\nu} = \int_{y_{0}}^{y_{max}} \int_{0}^{w(y)} \frac{2\, dw\, dy}{d^{2}} \frac{a_{j}}{a_{k}}\, \nu^{2}\, \int_{0}^{\tau_{\nu}} T(s) (e^{-(\tau^{ff}_{\nu} - \tau (s)) })\, d\tau
\end{equation}

\noindent Here, $y_{max}$ is the full extent of the jet. For low values of base density, $n_0$, the jet can be optically thin throughout its extent.


\begin{figure}
    \hskip -0.4cm
    \includegraphics[width=\columnwidth]{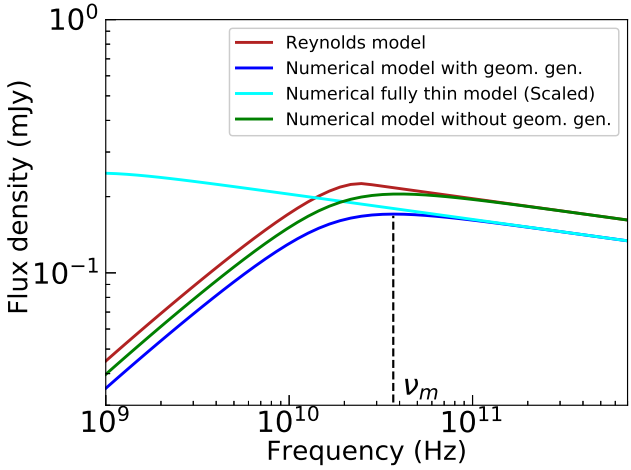}
    \caption{Comparison of thermal free-free spectrum calculated analytically (Reynolds, 1986) with that calculated numerically for a jet at a distance of 1~kpc. The parameters of the jet are $n_0$ = $5\times 10^7$~cm$^{-3}$, $x_0$ = 0.2, $q_n$ = -2, $q_x$ = 0, r$_0$ = 10~au, $\theta_0$ = 30$\degr$, $\epsilon$ = 1, $i$ = 60$^\circ$, T$_0$ = 10$^4$~K,  $q_T$= 0, $y_{max}$ = 1000~au, see text for more details. A lower $n_{0}$ of 10$^6$~cm$^{-3}$ results in a fully optically thin jet shown in cyan (scaled up by a factor of 2500). }
    \label{fig:compReynolds}
\end{figure}

An additional consideration that we include is the effect of inclination on the thickness of the jet. The inclination of the jet with respect to the sky plane will give rise to the jet having unequal LOS distances on front ($s_1 (y)$) and rear ($s_2 (y)$) sides of the long central jet axis. Thus the total LOS is $\mathscr{S} = s_1(y)\,+\,s_2(y)$ (as shown in Fig.~\ref{fig:geom}), rather than $2w/\sin{i}$ as assumed in the Reynolds model. For any $y$, $s_1(y)$ and $s_2(y)$ can be calculated from the jet geometry as follows (see Appendix~\ref{ap:varyingth} for details).
\begin{equation}\label{eq:s1}
s_{1}(y) = w(y)\, \frac{\cos{(\theta/2)}}{\sin{(i-\theta/2)}},\,\,\,\,s_{2}(y) = w(y)\, \frac{\cos{(\theta/2)}}{\sin{(i+\theta/2)}}
\end{equation} 
Note that these equations correspond to a specific case of a jet with constant opening angle, displayed in Fig.~\ref{fig:geom}. In general, at a particular $y$ for a jet with any $\epsilon$, these distances $s_1(y)$ and $s_2(y)$ can be calculated directly from the jet geometry using $\theta_{1}(y)$ and $\theta_{2}(y)$, respectively, which are the jet opening angles corresponding to $s_1(y)$ and $s_2(y)$ as shown in Fig.~\ref{fig:vary_th} in Appendix~\ref{ap:varyingth}.

\par The total flux density is calculated by taking this geometrical factor into consideration. The radio spectrum of the current model which incorporates intermediate ranges of optical depth is compared with that generated using Reynolds model. These are displayed in Fig.~\ref{fig:compReynolds}. The model parameters used to generate the spectra are $r_0 = 10$~au, opening angle $\theta_{0} = 30^{\circ}$, a typical jet base density $n_{0} = 5\times10^{7}$~cm$^{-3}$ which is estimated by the interpolation of typical values found in literature \citep{1995ApJ...449..184M, 1999A&A...342..717B} with a power-law index of $q_n$ = -2. The ionization fraction at the base of jet is $x_0 = 0.2$ \citep{1999A&A...342..717B}, temperature at base of jet is $T_{0} = 10^{4}$~K, constant opening angle $\epsilon = 1$, the jet is isothermal $q_T = 0$, constant ionization fraction $q_x = 0$, extent of the jet $y_{max} = 1000$~au, for inclination $i = 60^\circ$, at a distance $d = 1$~kpc. 
The frequency $\nu_m$ ($35.8$~GHz, blue curve) in the figure represents the turnover frequency where the full jet becomes optically thin ($\tau \le 1$).
 Hence, for frequencies below $\nu_m$, the spectral index is $+0.6$ as expected and above $\nu_m$, the spectral index is $-0.1$ which is characteristic of optically thin free-free emission. In addition to this, we have calculated the spectrum of a fully optically thin jet (cyan), which is attained at lower number densities, $n_{0}$ = $10^6$~cm$^{-3}$, with the other parameters remaining the same. In the figure, for the purpose of comparison, the resulting flux densities for this case are scaled up by a factor of $2500$, as the flux density is proportional to the square of number density. We have also compared with a model that is identical to the Reynolds model but includes the intermediate optical depth effects (green), i.e. geometric generalization is not included. As expected, we find a flux difference of $11-13\%$ \citep{2018A&ARv..26....3A}. The spectra in Fig.~\ref{fig:compReynolds} show a single turnover in the frequency range of interest. However, in general, a thermal jet could exhibit two turnover frequencies which separates a low frequency regime having a spectral index of $+2$ (fully optically thick), an intermediate frequency regime with spectral index close to $+0.6$ (part of the jet is optically thick and part is optically thin) depending on the geometrical parameters, and a high frequency regime with a spectral index of $-0.1$ (fully optically thin). The high frequency turnover is dictated by the ionized density at $ r_0$, while the low frequency turnover depends on the density close to $y_{\text{LOS}_{\text{max}}}$.

For our numerical model, the overall behavior is similar to the spectrum from Reynolds model. However, the Reynolds approximation is an overestimate by $24-32\%$, as expected. There are two factors that contribute to this discrepancy, which are as follows: (i) geometrical effects due to the large opening angle, and (ii) effect of intermediate optical depths. In our model each slice of the jet at any $y$ has an elliptical or circular cross-section depending on the jet inclination. In Reynolds model, on the other hand, a similar slice of the jet has rectangular cross-section. Due to the accommodation of larger opening angles in our model, the calculation of radial variation of the parameters are considered for a given LOS whereas the same has been ignored in the Reynolds model. The secondary factor for the difference in flux density is due to the inclusion of intermediate optical depths in the flux density estimation. The effect of intermediate optical depths is visible from the figure, where our model shows a relatively gradual turnover unlike the Reynolds case.


\begin{figure}
	\hskip -0.4cm
	\includegraphics[width=\columnwidth]{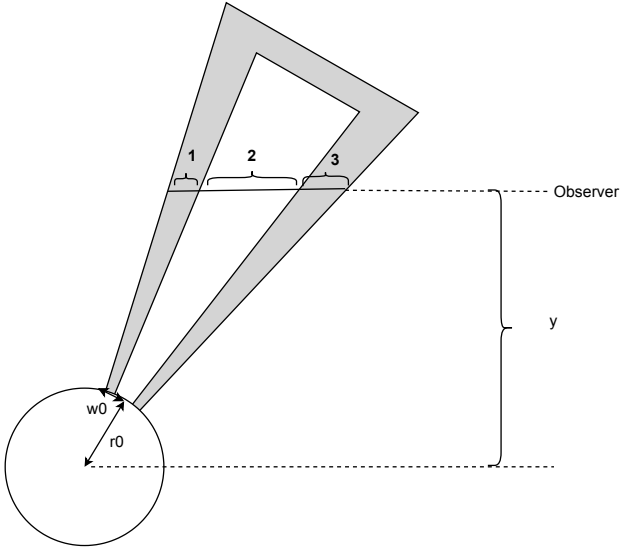}
    \caption{Schematic diagram of the jet representing the different layers of particle distribution along a particular LOS at $y$. Regions~1 and 3 give rise to a combination of thermal and non-thermal emission, while Region 2 solely contributes to thermal emission.}
    \label{fig:jet_layers}
\end{figure} 

\subsection{Combination of Free-Free and Synchrotron emission} \label{sync}

\par It has been observed that few protostellar jets display negative spectral indices steeper than $-0.1$ across radio frequencies \citep{1993ApJ...416..208M,1993ApJ...415..191C,1996ApJ...459..193G,1999ApJ...513..775W,2016ApJ...818...27R,2016MNRAS.460.1039P,2018MNRAS.474.3808V} indicating the presence of synchrotron emission. In few cases, towards higher frequencies, the spectral indices are flatter than what is expected from synchrotron emission \citep{2018MNRAS.474.3808V,1993ApJ...416..208M} suggesting the existence of non-thermal synchrotron emission along with thermal free-free emission. The generation of relativistic electrons that are accelerated back and forth at the shocks in the presence of magnetic field provides suitable conditions for the generation of synchrotron radiation \citep{1949PhRv...75.1169F,1987PhR...154....1B}. 

Following the confirmation of synchrotron emission through polarized emission towards HH~80-81 by \citet{2010Sci...330.1209C}, high resolution radio imaging and spectral index analyses by \citet{2017ApJ...851...16R} and \citet{2016ApJ...818...27R} have revealed that non-thermal emission is produced at the edges of the jet. Motivated by this, we explore the possibility of combining thermal and non-thermal emission from protostellar jets and we model it using this consideration.

\par We have already discussed the case of thermal free-free emission in Sect.~\ref{ff}, hence we discuss the case of synchrotron emission here. For a relativistic electron of mass $m$ and charge $e$, traveling at velocity $\vec{v}$ in a magnetic field $\vec{B}$, a Lorentz force $\vec{v} \times \vec{B}$ is exerted on the electron. This force accelerates the electron resulting in a radiative power loss which in turn depends on the electron energy and magnetic field strength. For a relativistic electron emitting synchrotron radiation at a characteristic frequency of $\nu_{pk}$, the corresponding Lorentz factor $\gamma_{pk}$ is given by the following equation \citep{rybicki2008radiative}.
\begin{equation}\label{eq:Lorentzf}
\gamma_{pk} = \sqrt{\frac{4\pi\, m\, c\, \nu_{pk}}{3\, e\, B} }
\end{equation}
\noindent This assumes an equi-partition of total energy into energy of relativistic electrons and energy due to the magnetic field (of magnitude $B$) in the shocked plasma in order to achieve a condition of minimum energy state \citep{2004IJMPD..13.1549G,1970ranp.book.....P}.
We consider a population of relativistic electrons in the shocked region, having a power-law number density distribution with an index $p$ in energy space, or the equivalent $\gamma$-space where $\gamma$ is the Lorentz factor. The kinetic energy $E$ of the electrons is related to their Lorentz factor as $E = \gamma\, m\, c^2$.

The number density of relativistic electrons, $n(\gamma)d\gamma$ between $\gamma$ and $\gamma+d\gamma$ is, therefore, given by the following equation.
\begin{equation}\label{eq:num_dist}
n(\gamma)\,d\gamma = n_{k}\,\gamma^{-p}\,d\gamma
\end{equation}

\noindent Here $n_k$ represents the proportionality constant for the number density of electrons in $\gamma$-space. For the given distribution of electrons, the synchrotron emission coefficient, $j^{syn}_{\nu}$, is given by the following expression \citep{rybicki2008radiative}.

\begin{equation} \label{eq:emission}
\begin{split}
j^{syn}_{\nu}& = \frac{1}{4\pi} \frac{\sqrt{3}\, e^{3}\, n_k\, B\, \sin{\alpha_\text{pa}} }{2\pi\, m\, c^{2}\, (1+p)}\, \Gamma\left(\frac{p}{4} + \frac{19}{12}\right)\, \Gamma\left(\frac{p}{4} - \frac{1}{12}\right) \times\\
 &  \bigg(\frac{2\pi\, \nu\, m\, c}{3e\, B\, \sin{\alpha_\text{pa}}}\bigg)^{-(p-1)/2} 
\end{split}
\end{equation}
Here $\Gamma(x)$ is the gamma function for an argument $x$ and $\alpha_\text{pa}$, termed as pitch angle, is the angle between the magnetic field and the velocity of electrons accelerated in the field. The synchrotron emission spectrum peaks at a critical frequency ($\nu_{pk}$) which corresponds to the turnover frequency at which the electrons emit the maximum power. 

\par It is also possible for the same population of relativistic electrons to absorb the synchrotron emission. This is expressed through the absorption coefficient, $\alpha^{syn}_{\nu}$, which is given by the following expression. 

\begin{flushleft}
\begin{equation} \label{eq:absorption}
\begin{split}
\alpha^{syn}_{\nu} &= \frac{\sqrt{3} e^{3}}{8\pi\, m}\, \bigg(\frac{3e}{2\pi m^{3} c^{5}}\bigg)^{p/2}\, n_k\, (B\, \sin{\alpha_\text{pa}})^{(p+2)/2}\, \Gamma \left(\frac{3p+2}{12}\right) \\
 & \times \Gamma \left(\frac{3p+22}{12}\right)\, \nu^{-(p+4)/2}
\end{split}
\end{equation}
\end{flushleft}
We assume an isotropic distribution of electron velocities. Therefore, the average of terms involving sin($\alpha_\text{pa}$) over all directions around the field is of the order of unity in Eqns.~\eqref{eq:emission} and \eqref{eq:absorption}.
The absorption coefficient is used to determine the optical depth, and the frequency where the synchrotron optical depth $\tau_{syn}=1$ defines the self-absorption frequency ($\nu_{a}$). This is determined by the microscopic parameters and the power-law index $p$. Depending on the parameters of the jet, $\nu_a$ can lie on either side of $\nu_{pk}$. Another feature of the spectrum is that, as the system ages, the energy of the electrons reduces due to radiative losses, resulting in lowering of Lorentz factors of the emitting electrons. With time, electrons with higher energies lose their energy faster, also known as spectral aging. Consequently, the higher energy tail of the spectrum breaks at a frequency (cooling frequency, $\nu_c$) where the cooling of those electrons becomes dominant. Thus, $\nu_c$ can be taken to be the upper frequency limit for our analysis as it is relatively large $\nu_c=100$~GHz \citep{2004IJMPD..13.1549G}.

For a distribution of electrons of density $n(\gamma)$ emitting between frequencies $\nu_{pk}$ and $\nu_c$, the constant $n_k$ in Eqn.~\eqref{eq:num_dist} can be estimated by integrating $n(\gamma)d\gamma$ from $\gamma_{pk}$ to $\gamma_c$. This results in the following expression.
\begin{equation}\label{eq:n_const}
n_k = \frac{n\, (p-1)}{\gamma_{pk}^{-(p-1)}}
\end{equation}


\begin{figure}
	\hskip -0.4cm
	\includegraphics[width=\columnwidth]{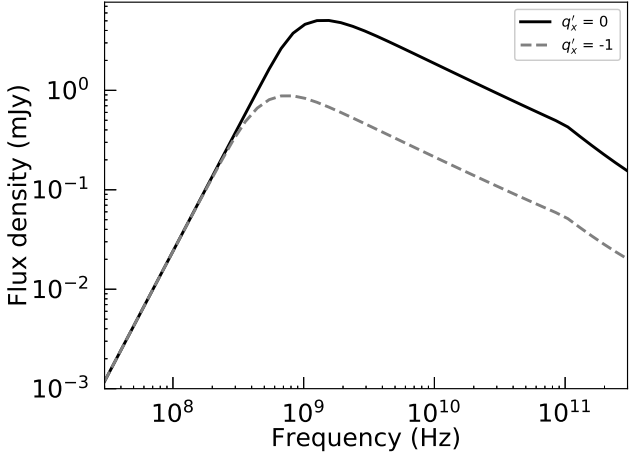}
    \caption{The spectra of jet models which incorporate both free-free and synchrotron emission. The parameters of the jet models are $n_0$ = 500~cm$^{-3}$, $q_n$ = -2, $T_0$ = 10$^4$~K, $r_0$ = $3000$~au, $\theta_0$ = 30$^\circ$, $x_0$ = 0.2, y$_{max}$ = $4500$~au, $\epsilon$ = 1, $i$ = 60$^\circ$ and $d$ = 1~kpc. The synchrotron radiation is assumed to be generated from a shell of thickness 0.5$^\circ$ having an electron distribution with $p = 2.3$ accelerated in a $\vec{B}$ field of 0.3mG, generating $\eta^{rel}_{e} = 10^{-5}$ in both the cases shown. The power-law index of radial variation in ionization fraction $q_x$ = -0.5. The power-law index of the lateral variation in ionization fraction is represented as $q'_x$; two cases of which are shown in the plot.}
    \label{fig:ff_nt_NOx}
\end{figure}

\par The calculation of the synchrotron spectrum is given in Appendix~\ref{ap:sync_spec}. We note that for a given distribution of electrons of constant number density, the spectral indices ($\beta$) across the spectrum due to synchrotron emission varies as follows:\\

For $\nu_a<\nu_{pk}\,:$ \\
\hspace*{1cm} $\nu<\nu_a$, $\beta = 2$\\
\hspace*{1cm} $\nu_a<\nu<\nu_{pk}$,  $\beta = 1/3$\\
\hspace*{1cm} $\nu_{pk}<\nu<\nu_c$, $\beta = -(p-1)/2$\\
\hspace*{1cm} $\nu>\nu_c$, $\beta = -p/2$\\

For $\nu_a>\nu_{pk}\,:$\\
\hspace*{1cm} $\nu<\nu_{pk}$, $\beta = 2$\\
\hspace*{1cm} $\nu_{pk}<\nu<\nu_a$,  $\beta = 5/2$\\
\hspace*{1cm} $\nu_a<\nu<\nu_c$, $\beta = -(p-1)/2$\\
\hspace*{1cm} $\nu>\nu_c$, $\beta = -p/2$\\

\noindent In our model we do not encounter a case of $\nu_a < \nu_{pk}$ due to the energies involved, hence we ignore this part of the spectrum calculation for the present. 

Armed with the knowledge about both, the free-free and synchrotron emission mechanisms discussed so far, we proceed towards the methods by which these mechanisms can be incorporated in the geometry to find the radio flux densities generated by the jet. For a given LOS, we can divide the jet into three regions as shown in Fig.~\ref{fig:jet_layers}. Regions 1 and 3 (also R1 and R3) represent the outer shocked material which are the front and rear layers, respectively, while viewing along $\mathscr{S}$. These regions contribute to the flux densities through (i) synchrotron radiation emitted by a fraction of electrons that are relativistic ($\eta^{rel}_{e}$), and (ii) free-free emission from the rest of the non-relativistic fraction of electrons. The lateral variation in ionization fraction which is introduced in Sect.~\ref{ionizefrac} are incorporated here in all the three regions R1, R2 and R3. As the distributions of populations generating (i) and (ii) are independent in a given volume, the overall emission and absorption coefficients for the combination of the processes are given by the individual sums of the coefficients of the two processes. These are given by the following expressions.

\begin{equation}\label{eq:jcoeff_coupled}
j^{ff+syn}_{\nu}(s,y) = j^{ff}_{\nu}(s,y) + j^{syn}_{\nu}(s,y)
\end{equation}
\begin{equation}\label{eq:alcoeff_coupled}
\alpha^{ff+syn}_{\nu}(s,y) = \alpha^{ff}_{\nu}(s,y) + \alpha^{syn}_{\nu}(s,y) 
\end{equation}

Region 2 (also R2) is the inner highly ionized jet (near the jet-axis) which solely emits free-free radiation, and therefore has emission and absorption coefficients $j^{ff}_{\nu}$ and $\alpha^{ff}_{\nu}$, as discussed earlier. Hence a LOS along $\mathscr{S}$ intersects with different particle distributions and the intensities contributed by each of these regions for a given $y$ are given by the following expressions.

\begin{equation}\label{eq:Iregion1}
	I^{ff+syn}_{\nu, R1} = \int^{\tau^{ff+syn}_{\nu, R1} }_{0} \frac{j^{ff+syn}_{\nu, R1}}{\alpha^{ff+syn}_{\nu,R1}}  e^{-(\tau^{ff+syn}_{\nu, R1} - \tau)}\,d\tau
	\end{equation}
	
	\begin{equation}\label{eq:Iregion2}
	I^{ff}_{\nu, R2} = \int^{\tau^{ff}_{\nu,R2}}_{0}  \frac{j^{ff}_{\nu,R2}}{\alpha^{ff}_{\nu,R2}}\,e^{-(\tau^{ff}_{\nu,R2}- \tau)}\,d\tau
	\end{equation}

	\begin{equation}\label{eq:Iregion3}
	I^{ff+syn}_{\nu, R3} = \int^{\tau^{ff+syn}_{\nu,R3}}_{0} \frac{j^{ff+syn}_{\nu,R3}}{\alpha^{ff+syn}_{\nu,R3}} e^{-(\tau^{ff+syn}_{\nu,R3} - \tau )}\,d\tau
	\end{equation}

Here, $I_{\nu,Rn}$ and $\tau_{\nu, Rn}$ represent the total intensity and optical depth for each region given by subscript $n$, where $n = 1,\,2,\,3$. Note that the limits of the integration are dependent on the region under consideration and the projected length $y$. The total radiation intensity from a LOS at $y$ can be calculated by adding the contribution of each of these regions after considering absorption from the layers in front of it. This is given as follows.
\begin{equation}\label{eq:total_intensityI}
I_{\nu}(y) = I^{ff+syn}_{\nu,R1}\,e^{-(\tau^{ff}_{R2}\,+\, \tau^{ff+syn}_{R3})}\,+\, I^{ff}_{\nu,R2}\,e^{-\tau^{ff+syn}_{R3}} +\, I^{ff+syn}_{\nu,R3}
\end{equation}
The above intensity is integrated along the width (in the sky plane) and the projected length of the jet to give the total flux density emitted by the jet.
\begin{equation}\label{eq:tot_wy}
S_{\nu} = \int_{y_{0}}^{y_{max}} \int_{0}^{w(y)} \frac{2\, dw\, dy}{d^{2}} I_{\nu}(y)
\end{equation}

\noindent It should be noted that $I_{\nu}(y)\equiv I_{\nu}(w,y)$. This is because so far we have discussed only about the LOS which crosses the main axis of the jet but, the LOS through the jet decreases as one moves from the axis along the width of the jet in the plane of the sky. This discussion can be found in Appendix~\ref{ap:shell}. In addition to the inner region and outer shell, we also consider the interaction of the jet with the ambient medium at the terminal region and for simplicity assume a flat thin interaction region instead of bow-shock like curvature seen towards terminal edges of many jets and HH objects. This region produces synchrotron emission, in addition to thermal emission. The LOS through the extreme end of the jet, therefore, gradually tapers down due to its inclination with respect to the plane of the sky. Hence, at the extreme top end of the jet, only the shocked region contributes to the total flux, identical to R1 and R3.

\par An example of a jet spectrum with thermal and non-thermal emission is shown by the solid black curve in Fig.~\ref{fig:ff_nt_NOx}. Here, we have assumed the jet to have a constant opening angle of $\theta_0=30^\circ$, of which the outer region $\delta\theta=0.5^\circ$ contributes to the synchrotron emission, and for this we have taken $\eta^{rel}_{e}= 10^{-5}$. Due to episodic ejection of materials by the driving source, jets are usually observed as knots/lobes. Since non-thermal emission is seen in jet knots which are observed to be farther away from the central source, we have assumed an $r_0$ = $3000$~au and the size of the lobe to be $800$~au. We have considered an electron distribution with $p = 2.3$ accelerated in a $\vec{B}$ field of $0.3$~mG. The base number density is taken as $n_0 = 500$~cm$^{-3}$ with number density falling off as $r^{-2}$, and the jet is isothermal with a temperature $T=10^4$~K. The ionization fraction at the jet base $x_0 = 0.2$ with its radial variation given by $q_x = -0.5$. We have considered a value of $\nu_{pk}=10$~MHz and $\nu_c=100$~GHz. The other parameters are the same as those used earlier to generate the free-free emission in Fig.~\ref{fig:compReynolds}. The radio spectrum seen in Fig.~\ref{fig:ff_nt_NOx} (solid black curve) displays dominant synchrotron contribution, with a turnover at $\nu_t=1.4$~GHz. This turnover represents the frequency at which the flux density is maximum and is a sort of average over the entire jet, determined by the combined properties of free-free and synchrotron emission. The spectral indices are $+2.5$ below $\nu_t$, and $-0.64$ above $\nu_t$. There is a break in the spectrum at higher frequency that occurs at $\nu_c$, corresponding to the synchrotron cooling frequency as discussed earlier.

\par  In general, one could encounter different types of turnovers in a spectrum that incorporates both thermal free-free and synchrotron emission. In frequency ranges where either thermal or non-thermal emission mechanism dominates, there could be frequency turnovers corresponding to a fully optically thick jet to a regime where part of the jet is optically thick and the rest is optically thin (change of spectral index from 2 to $\sim$0.6), from this regime to a fully thin thermal jet ($\sim$0.6 to -0.1), and from optically thick to thin regime for the non-thermal emission (2.5 to -0.6). In addition, the combination of the thermal and non-thermal contributions can result in more turnovers depending on the frequency sub-ranges in which alternate emission mechanisms dominate. This could include the transition from fully optically thick thermal emission to optically thin non-thermal emission (2 to -0.6), from fully optically thick thermal emission to optically thick non-thermal emission (2 to 2.5), from optically thick non-thermal emission to a regime where a part of the jet is optically thick and part is optically thin thermal emission (2.5 to $\sim$0.6), from optically thin non-thermal emission to fully thin thermal emission (-0.6 to -0.1), etc. It is important to bear in mind that the above mentioned spectral indices are indicative and the exact values will depend on the relative contributions of the emission from the two mechanisms.


\begin{figure}
	\hskip -0.4cm
	\includegraphics[width=\columnwidth]{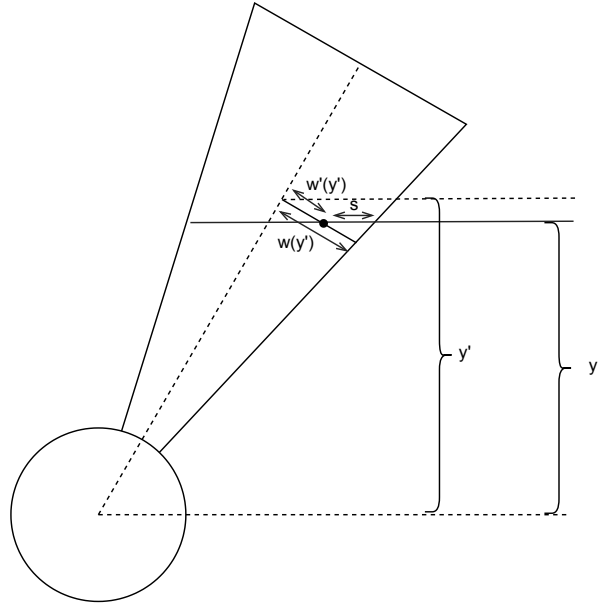}
    \caption{Schematic diagram of the jet representing the distance of any point $s$ along the LOS from the long axis [$w'(y')$], and the corresponding jet width [$w(y')$].}
    \label{fig:ionization_fraction}
\end{figure}


\begin{figure*}
\minipage{\textwidth}
 \includegraphics[width = 0.5\columnwidth]{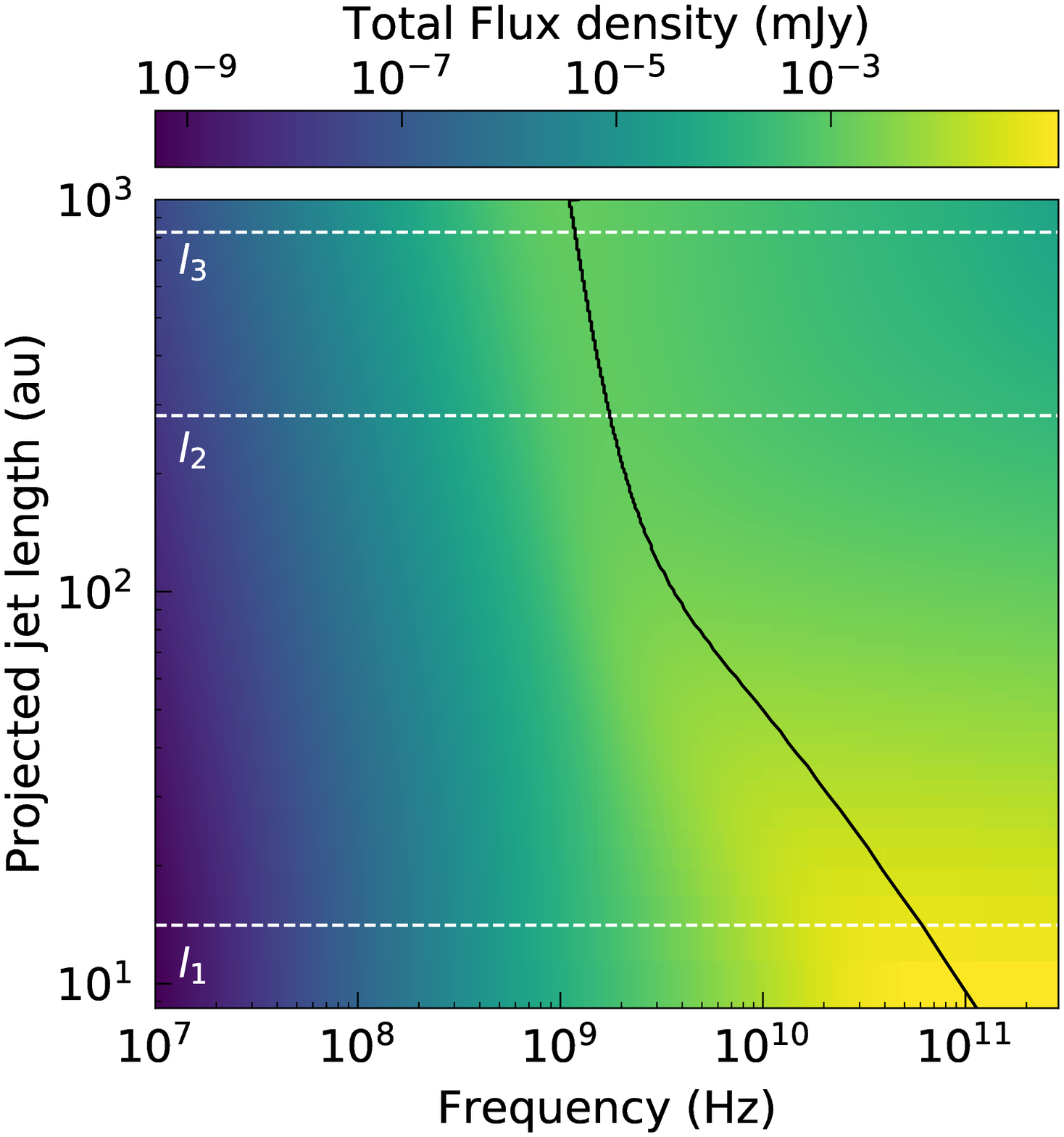}
\endminipage
\minipage{\textwidth}
\vspace*{1cm}
\hspace*{-8cm}
  \includegraphics[width = 0.45\columnwidth]{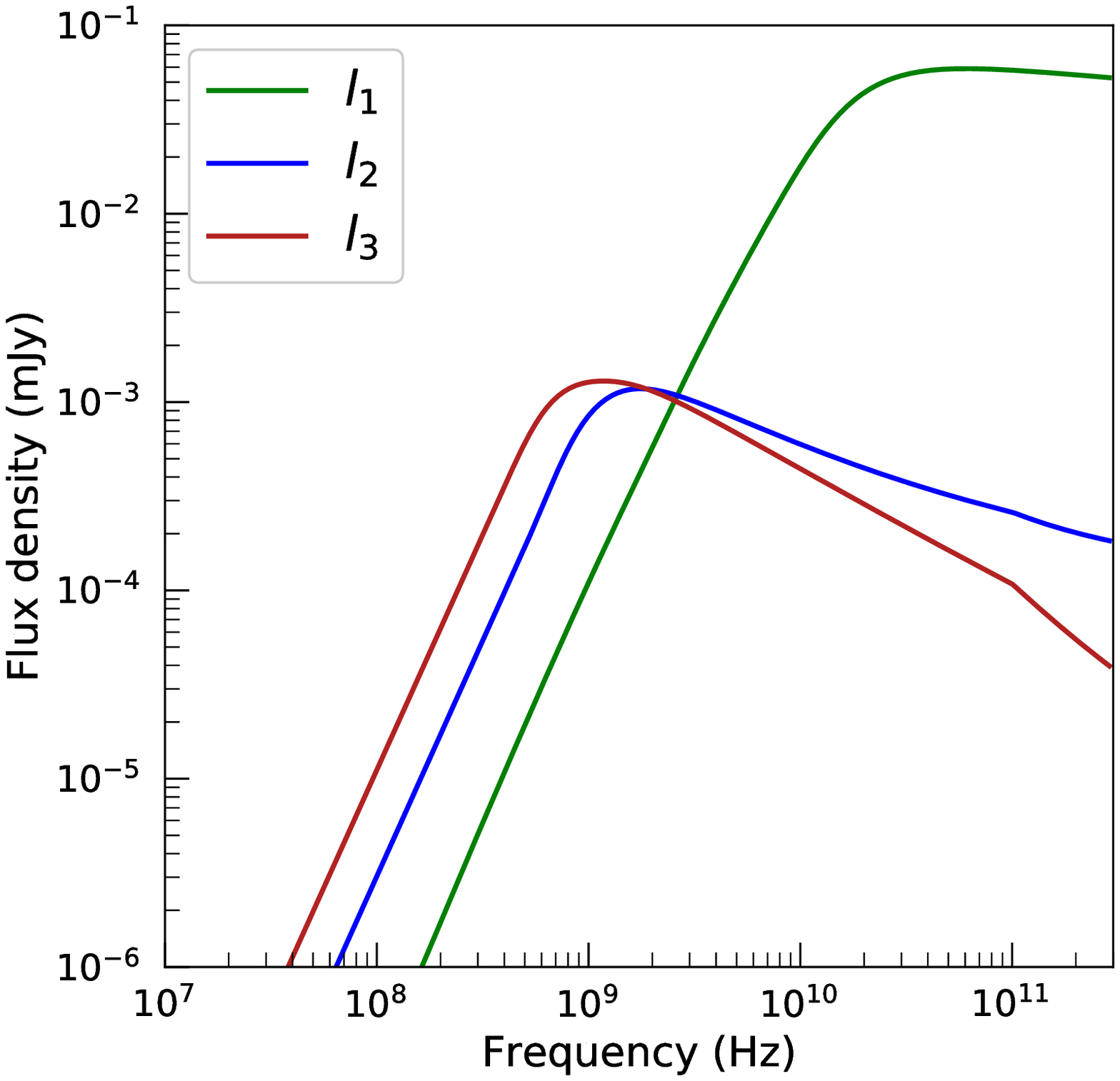}
\endminipage
\caption{[Left] The flux density distribution (combined free-free and synchrotron) as a function of frequency ($\nu$) and projected length of the jet ($y$) plotted in logarithmic space. At each $y$ along the jet, the flux densities are calculated for segments with width equal to the width of the jet, and $\Delta y$ = 3~au. The black curve represents the variation of the turnover frequency $\nu_t$ where emission is maximum. $l_1$, $l_2$ and $l_3$ marked in white are three different positions along $y$, whose spectra are shown in [right]. The jet model parameters $n_0$ = 10$^8$~cm$^{-3}$, $r_0$ = 10~au, $y_{max} = 1000$~au, $q_x$ = $q'_x$ = 0; the remaining paramaters are same as in Fig.~\ref{fig:ff_nt_NOx}.}
\label{fig:colormap_spec}
\end{figure*}

\section{Lateral variation of ionization fraction} \label{ionizefrac}

\par The ionization fraction or the degree of ionization plays an important role in quantifying the coupling of the gas to the magnetic field. The distribution of ionized gas in the jet is important as theories predict that the centrifugal launch of jet occurs along the magnetic field lines \citep{1999ApJ...524..159G,2014EPJWC..6405005Z}. 

Observationally, studies of ionized jets from massive YSOs have indicated ionization fraction to be typically $2-15\%$ with the maximum value measured to be upto $40\%$ \citep{{2005A&A...441..159N}, {2002RMxAC..13....8B},{1999A&A...342..717B}, {Giannini_2013}, {2019MNRAS.486.3664O}}. In addition, it has been observed that ionization fractions decrease with radial distance away from the central source as expected \citep{2019NatCo..10.3630F}, lending credence to the introduction of the parameter $q_x$ in the models, which has the following form.
\begin{equation}\label{eq:x_axial}
x_a(r) = x_0 \left(\frac{r_{a}}{r_0}\right)^{q_x}\,\,\,\Rightarrow\,\,\,\,\, x_a(y) = x_0 \left(\frac{y}{y_0}\right)^{q_x} 
\end{equation}
\noindent Here, the subscript $a$ implies that it is axial and $x_0$ represents the base ionization fraction at $r_0$. 

The model discussed so far allows for a variation of ionization fraction along the radial direction only, through $q_x$. We believe it is realistic to also incorporate an ionization fraction that decreases across the jet cross-section, away from the central long axis towards the edge of the jet. This is expected because theories predict a decrease in the velocity structure laterally across jets \citep{2014ApJ...796L..17M}. This suggests that the inner regions of the jet would be highly ionized and the ionization fraction decreases to 0 towards the lateral edges. The presence of synchrotron emission from shocks at the edges of jets is another inducement for introducing the lateral variation in ionization fraction as it has been found from line ratios investigated in shocks from jets that the ionization fractions are found to be relatively low, $\sim1-10\%$ \citep{1994ApJ...436..125H,1995A&A...296..185B,2010A&A...521A...7D}.

\par In order to establish the effect of ionization fraction across the lateral direction of the jet, we consider the following modifications to the equations discussed in Sect.~\ref{ff} and \ref{sync}. The variation of ionization fraction across a given LOS of the jet can be introduced through a power-law with index $q_x'$ as follows.
\begin{equation}\label{eq:x}
x(s) = x_a(y')\left[\frac{w(y')}{w(y') - w'(y')}\right]^{q_x'}
\end{equation}
\noindent Here, $y'$ is a function of $s$, $x_a(y')$ is the ionization fraction along the long axis of the jet at $y'$, $w'(y')$ represents the shortest distance of any point $s$ along a given LOS, from the central long axis of the jet, while $w(y')$ is the half-width of the jet that intersects the point $s$, shown schematically for clarity in Fig.~\ref{fig:ionization_fraction}. We consider $q_x'\leq0$ as ionization diminishes towards the edges. 
For a jet with $q_x=0$ and $q_x'=0$ we take $x(s)=1$ throughout, even though a singularity appears at the outer edges, i.e. at $w=w'$. 

\par These modifications are incorporated to calculate the number density of electrons in Regions 1, 2 and 3, and more details are presented in Appendix~\ref{ap:shell}. The new spectrum generated with $q'_x < 0$ is expected to have lower flux densities compared to the case of $q'_x = 0$. This is evident from the spectrum generated for  $q'_x = -1$ (grey dashed curve), and shown in Fig.~\ref{fig:ff_nt_NOx} for comparison. All the other parameters are kept the same as that for the solid black curve described in Sect.~\ref{sync}. The turnover frequency in this case $\nu_t = 750$~MHz, which is lower than that for $q'_x = 0$.

\par  It is instructive to note the variation in flux densities across the length of the protostellar jet. For the model parameters considered above and $q_x'= -1$, we present the total flux densities as a function of the projected length along the y-axis and frequency along the x-axis, in Fig.~\ref{fig:colormap_spec}~(left). For this, the jet is divided into segments with width equal to the width of the jet at any $y$ and height $\Delta y = 3$~au. The flux densities from each of these segments is then calculated separately. The jet spectrum at any arbitrary $y$ can be analyzed by noting the behavior across the frequency axis. $l_1$, $l_2$ and $l_3$ marked in white represent three different locations along the jet projected length $y$. The spectra for the three different positions are shown in Fig.~\ref{fig:colormap_spec}~(right). Blue and green in Fig.~\ref{fig:colormap_spec}~(right) show that for $l_1$ and $l_2$, the total flux is dominated by free-free emission. Here, the major contribution of the flux density is from the inner thermal jet. On the other hand as one moves farther out to larger $y$, the flux contribution is dominated by synchrotron (red). Whether the dominant flux contribution to a spectrum is from thermal free-free or non-thermal synchrotron emission can be established from the spectral indices. In the sample spectra shown in Fig.~\ref{fig:colormap_spec}~(right), the optically thin spectral index along $l_1$ is $-0.07$ implying dominant thermal free-free emission, and those along $l_2$ and $l_3$ are $-0.40$ and $-0.60$, respectively, upto $100$~GHz ($\nu_c$). From this, we conclude that non-thermal synchrotron emission is dominant in the higher frequency range along these two lines-of-sight.  

\par In each of these spectra, one can identify a turnover frequency that corresponds to the combined (free-free and synchrotron) emission of the jet element towards this position becoming fully optically thin. We observe that this shifts to lower frequencies as one moves away from the jet base. From Fig.~\ref{fig:colormap_spec}~(left), it can also be seen that the rate of change in turnover frequency with maximum emission (black) becomes lower in the synchrotron dominated region of the jet implying that the synchrotron turnover frequency is less sensitive to radial variation in parameters such as jet width $w(y)$, number density $n$ and ionization fraction $x$ as compared to free-free emission. This is easily understood since we know that synchrotron and free-free optical depths vary with the density of electrons as $\tau^{syn}_{\nu}\propto n\,x$ and $\tau^{ff}_{\nu}\propto n^{2}\,x^2$. As we move outwards to larger $y$, $w(y)$ increases but $n$ and $x$ decrease, thereby exhibiting more influence on the free-free optical depth and hence its turnover frequency. These spectra clearly exhibit the transition from free-free to synchrotron dominated flux as one moves from the jet base to the top of the jet.

\section{Dependence of radio spectrum on model parameters}\label{ap:spec_free}

The dependence of the model spectrum on various parameters is illustrated here, both for (i) free-free emission analogous to thermal jets, and (ii) a combination of free-free and synchrotron emission. We characterize the spectrum in terms of the turnover frequencies. The full spectrum is divided into frequency windows based on the turnover frequencies and the spectral indices are determined accordingly in these windows.

\subsection{Thermal jet}\label{sub:thermal}

\begin{figure*}       
    \begin{minipage}{\textwidth}                
        \includegraphics[width = 0.335\textwidth]{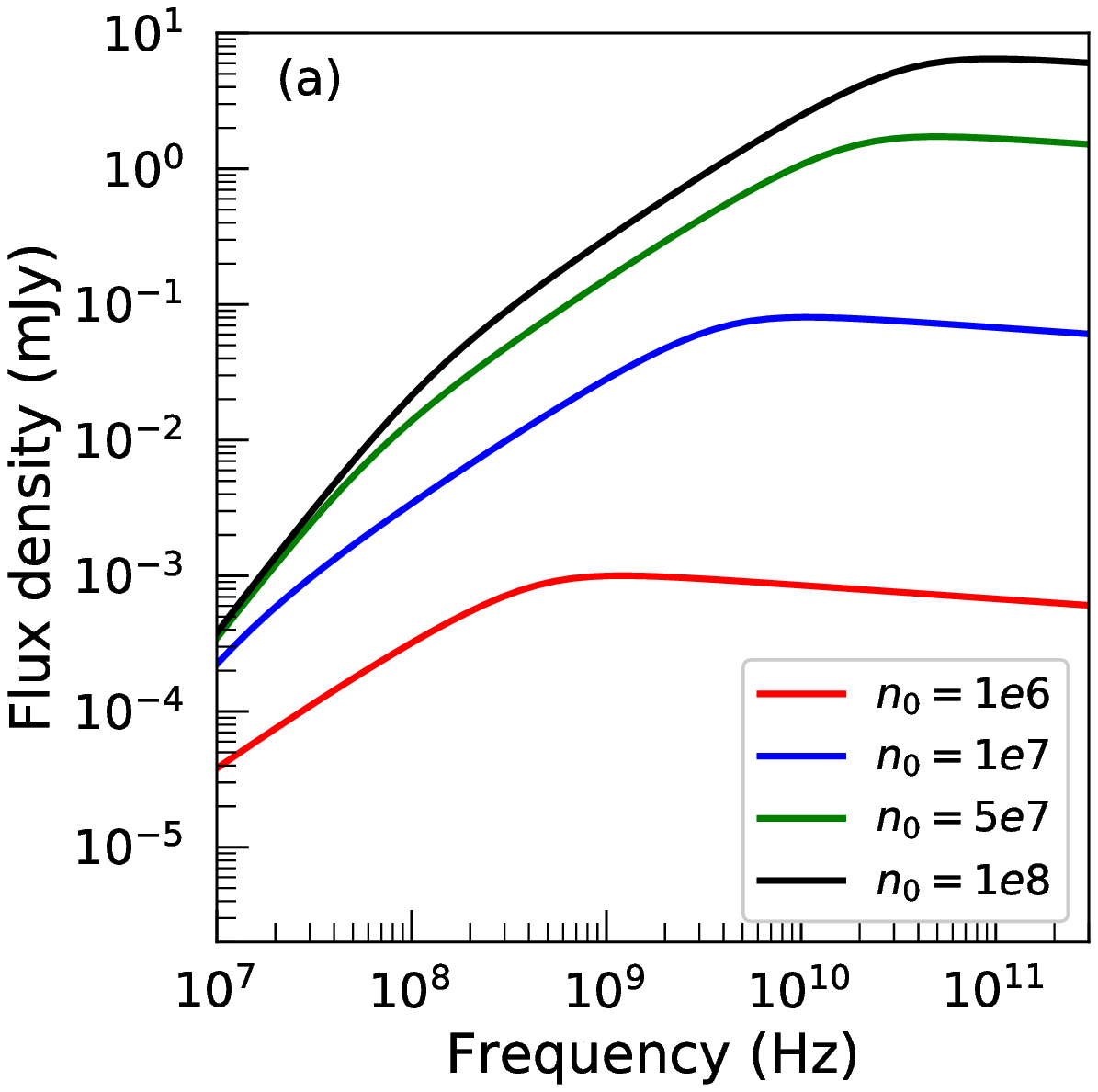}                 
        \includegraphics[width = 0.33\textwidth]{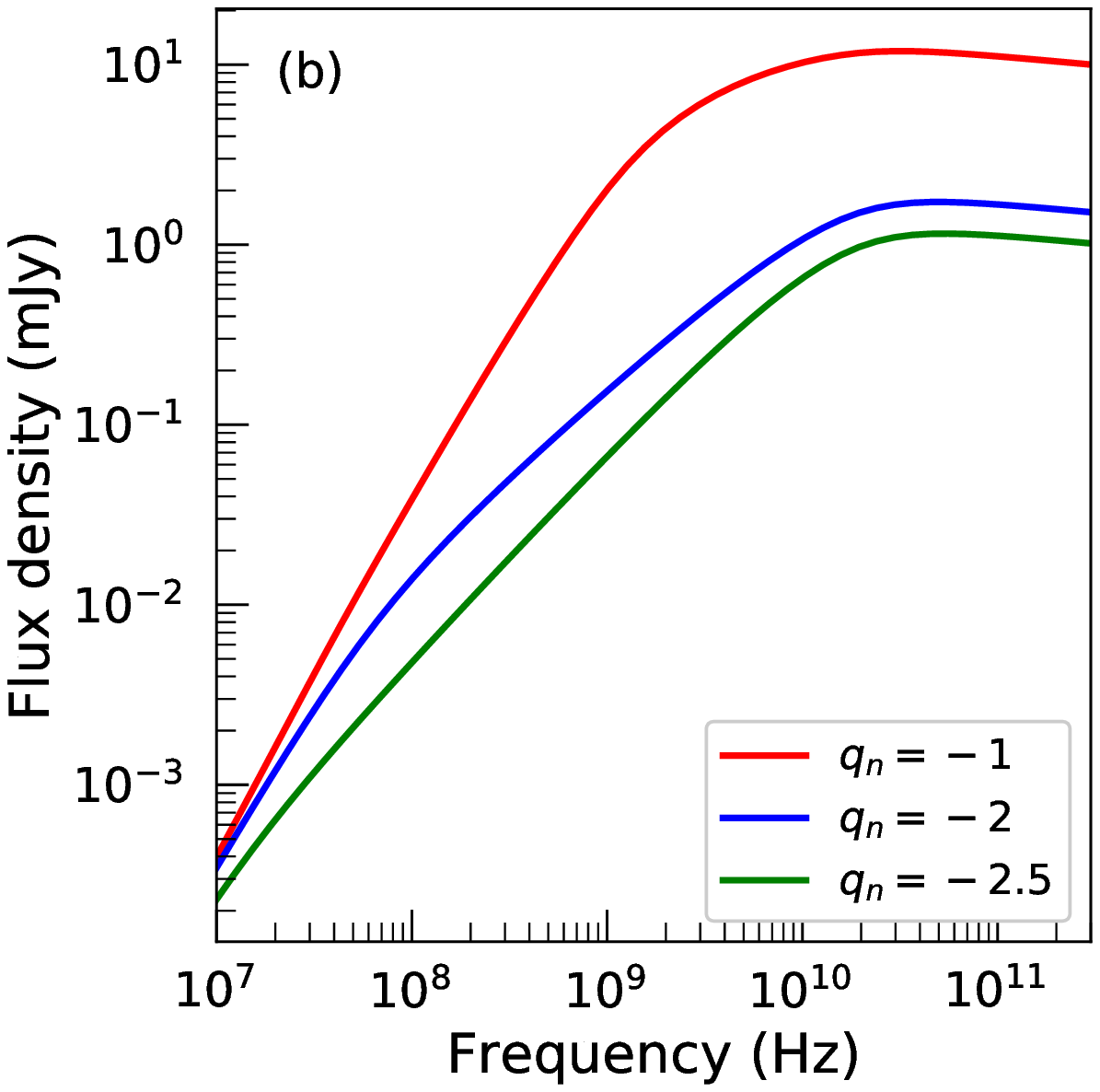}         
        \includegraphics[width = 0.33\textwidth]{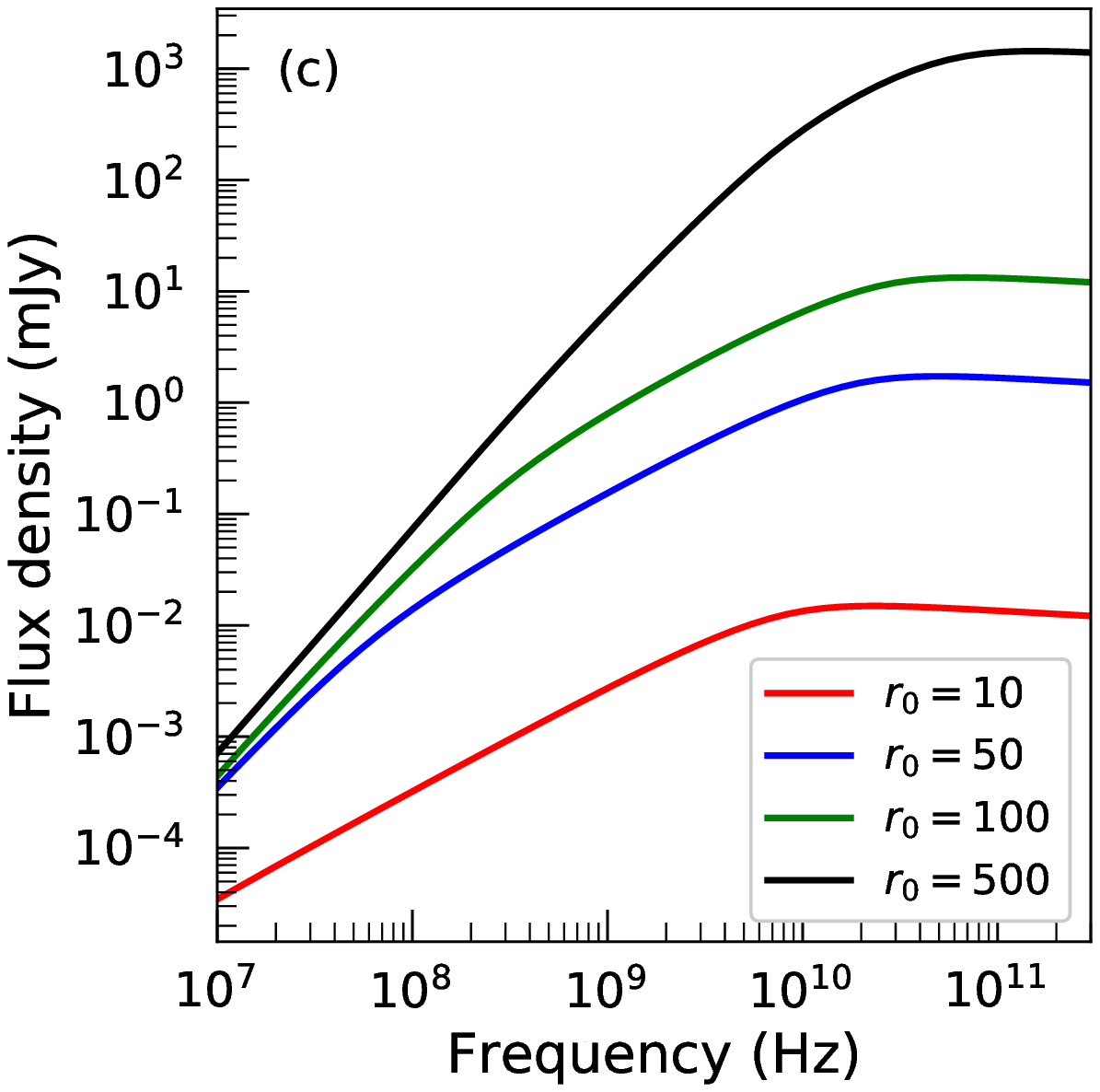} 
        \begin{center}       
        \caption{Dependence of the radio spectrum of a fully thermal jet on (a) $n_0$ $(cm^{-3}), $ (b) $q_n$, and (c) $r_0$~(au). Other than the parameter which is varied in each case, the remaining jet parameters are $r_0$ = 50~au, $n_0$ = $5\times 10^7$~cm$^{-3}$, $\theta_0 = 30^\circ$, $x_0$ = 0.2, $T_0$ = $10^{4}$~K, $q_n$ = -2, $q_x$ = -0.5, $q'_x$ = -1, $\epsilon$ = 1, $y_{max}\sim1300$~au, $i = 60^\circ$ and d = $1$~kpc. The details of the spectra shown in (a), (b) and (c) are listed in Table~\ref{tab:freeparms1} as Cases~I, II and III, respectively. } 
        \label{fig:set1}
         \end{center}
    \end{minipage}\hfill    
\end{figure*}

We first consider the cases of a fully thermal jet. For all the thermal spectra shown in this section, the high frequency turnover corresponds to the frequency above which $r_0$ becomes optically thin and the low frequency turnover corresponds to the frequency below which the emission towards $y_{\text{LOS}_{\text{max}}}$ becomes optically thick. Consider a jet located close to the YSO at $r_0 = 50$~au with $y_{max}\sim1300$~au, inclined at an angle $i = 60^\circ$, and located a distance $d = 1$~kpc. In this case, we do not include synchrotron emission. We assume the parameters at $r_0$ as follows: jet opening angle $\theta_0 = 30^\circ$, ionization fraction $x_0=0.2$ and electron temperature $T_0=10^4$~K. The remaining parameters are $\epsilon = 1$, the power-law indices of radial ionization fraction profile $q_x = -0.5$, and lateral ionization fraction profile $q'_x = -1$.  Fig.~\ref{fig:set1}~(a) shows the spectrum of this knot for various number densities $n_0$ at $r_0$, assuming $q_n = -2$~(Case~I). An increase in $n_0$ would lead to an overall increase in the number of particles contributing to the emission and consequently, the flux densities would increase. It can be seen from the figure that with an increase in number density by a certain factor, the flux density increases roughly by the same factor. It is instructive to compare our results with the predictions of the analytical Reynolds model, although the latter makes geometrical approximations and does not consider lateral variation in ionization fraction. We emphasize that the absolute values of flux densities in our model are expected to be different from that of Reynolds due to these generalizations.
The flux densities estimated from our model are about $65-86\%$ lower than the values from the corresponding Reynolds model. This can be attributed to contributions from (i) optical depth generalization, (ii) geometrical generalization that includes the (a) jet cross-section and (b) opening angle generalization, and (iii) introduction of lateral variation in ionization fraction. We find that a flux difference of about 24-32\% can be accounted due to (i) and (ii), while the rest is attributed to (iii). Of the 24-32\% flux difference, $\sim 8-12\%$ is due to the effect of intermediate optical depths, about 7-9\% is due to opening angle generalization, and $\sim 9-12\%$ is due to the modification of the jet cross-section (from rectangular cross-section in Reynolds model to circular cross-section in our model).
If we consider a relatively low frequency of 270~MHz (arbitrary), the ratio of flux densities for $n_0 = 10^7$, $5\times10^7$ and $10^8$~cm$^{-3}$ with respect to $n_0 = 10^6$~cm$^{-3}$ are 13, 62 and 115 in our model. According to the Reynolds model the flux densities are directly proportional to $n_0$, and hence the corresponding ratios are 10, 50 and 100. This implies the agreement between the two models at low frequencies. In Fig.~\ref{fig:set1}(a), we also note that as the number density increases, both the high and low frequency turnovers shift to higher values. With an increase in the number density of the particles contributing to the emission, the overall optical depth within the jet increases. As a result of this, the farther region of the jet with the largest LOS distance (that determines the low frequency turnover) will remain optically thick even at higher frequencies compared to the cases with lower number densities. Similarly, $r_0$ remains optically thick upto higher frequencies thereby resulting in an increase in the high frequency turnover. The Reynolds model predicts that the transition to a fully optically thin jet (high frequency turnover) is dependent on $n_0$ as $n_0^{2/2.1}$. This means that a variation of $n_0$ by a factor of 10 should result in a variation of turnover frequency by a factor of $\sim9$. The corresponding factor obtained from our model is $\sim10$. The dependence of turnover frequencies on $n_0$ can be observed from Table~\ref{tab:freeparms1}~(Case~I).

\par In Fig.~\ref{fig:set1}~(b), we demonstrate the effect of varying $q_n$ on the spectral properties for $n_0 = 5\times10^7$~cm$^{-3}$~(Case~II). The remaining parameters are the same as Case~I. The figure shows that the flux densities decrease with an increase in $|q_n|$. This is because a larger $|q_n|$ results in a lower number density of electrons as we move away from $r_0$ that contribute to the overall flux. To understand this better, we take the ratio of flux densities between $q_n$ = -2 and -2.5 at a frequency of 500~MHz suitable for comparison with the Reynolds model. The ratio is 2.6 for our model and the corresponding ratio for Reynolds  model is 2.9. The values are similar, but not identical. The difference can be attributed to geometrical generalization and lateral variation of ionization fraction included in our model. In the figure, the spectral indices in the frequency range between the two turnover frequencies are 0.36, 0.81 and 1.04 for $q_n$ = -1, -2 and -2.5, respectively. This implies that a larger $|q_n|$ results in a steeper spectrum. This is because an increase in $|q_n|$ results in a larger decline in overall jet density. As number density decreases, the portion of the jet that is optically thick gradually reduces and as a result, the flux contribution from optically thick region will be lower resulting in a steeper spectral index. The spectral indices and turnover frequencies corresponding to these plots are listed as Case~II in Table~\ref{tab:freeparms1}.

\par In this context, we consider the case of varying $r_0$~(Case~III). Fig.~\ref{fig:set1}~(c) shows the dependence of the spectrum on the parameter $r_0$ for a constant $n_0 =5\times10^7$~cm$^{-3}$. The remaining parameters are the same as Case~I. 
It is to be noted that this case represents spectra of different jets with different number densities although $n_0$ at $r_0$ for each jet is the same. This implies that a jet with larger $ r_0$ has a higher number density at any radial length $r$ as compared to the number density of a jet with a lower $r_0$. This means that the number density at each $r$ that is contributing to the emission will be different for each jet since the number densities are scaled by the factor $\big(\frac{r}{r_0}\big)^{q_n}$. This produces an indirect effect on the observed spectrum as $r_0$ is varied. To understand this in detail, we consider Eq.~(16) of the Reynolds model which states that the flux densities are proportional to $r_0^{2.5}$ for the assumed parameters. We again select a low frequency for comparison with the Reynolds model. At a frequency of 600~MHz, the ratio of flux densities for the cases of $r_0$ = 50 and 100~au with respect to $r_0 = 10$~au as obtained from the our model are 54 and 259, respectively, whereas the corresponding values predicted by the Reynolds model are 55 and 316. This conveys the agreement between the two models. For $r_0 = 500$~au, a comparison with the Reynolds model is difficult since this spectrum is dominated by fully optically thick and  fully optically thin regimes only, without a perceptible intermediate regime. This is because in this case, $r_0$ and $r_{max}$ are rather close, so that the low and high frequency turnovers are quite similar. We again find that for these parameters, our model flux densities are lower by $67-86\%$ as compared to the Reynolds model across the spectrum and we attribute this difference to the effect of geometrical generalization and introduction of lateral variation in ionization fraction. The turnover frequencies for this case are given in Table~\ref{tab:freeparms1}~(Case~III).


\begin{figure}  
    \begin{minipage}{0.5\textwidth}  
        \hspace*{-0.8cm}      
        \includegraphics[width = 0.55\textwidth]{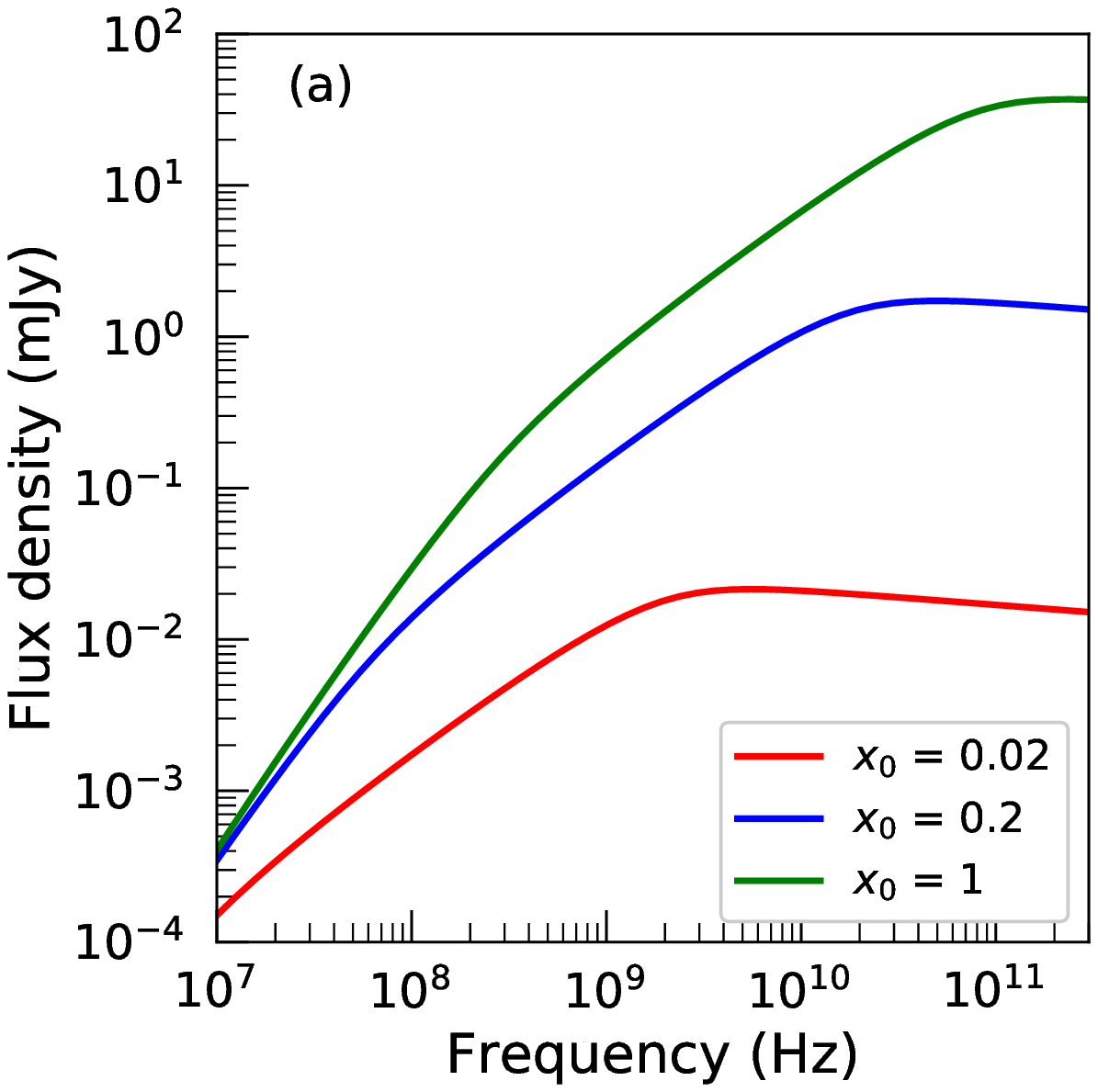}        
        \includegraphics[width = 0.54\textwidth]{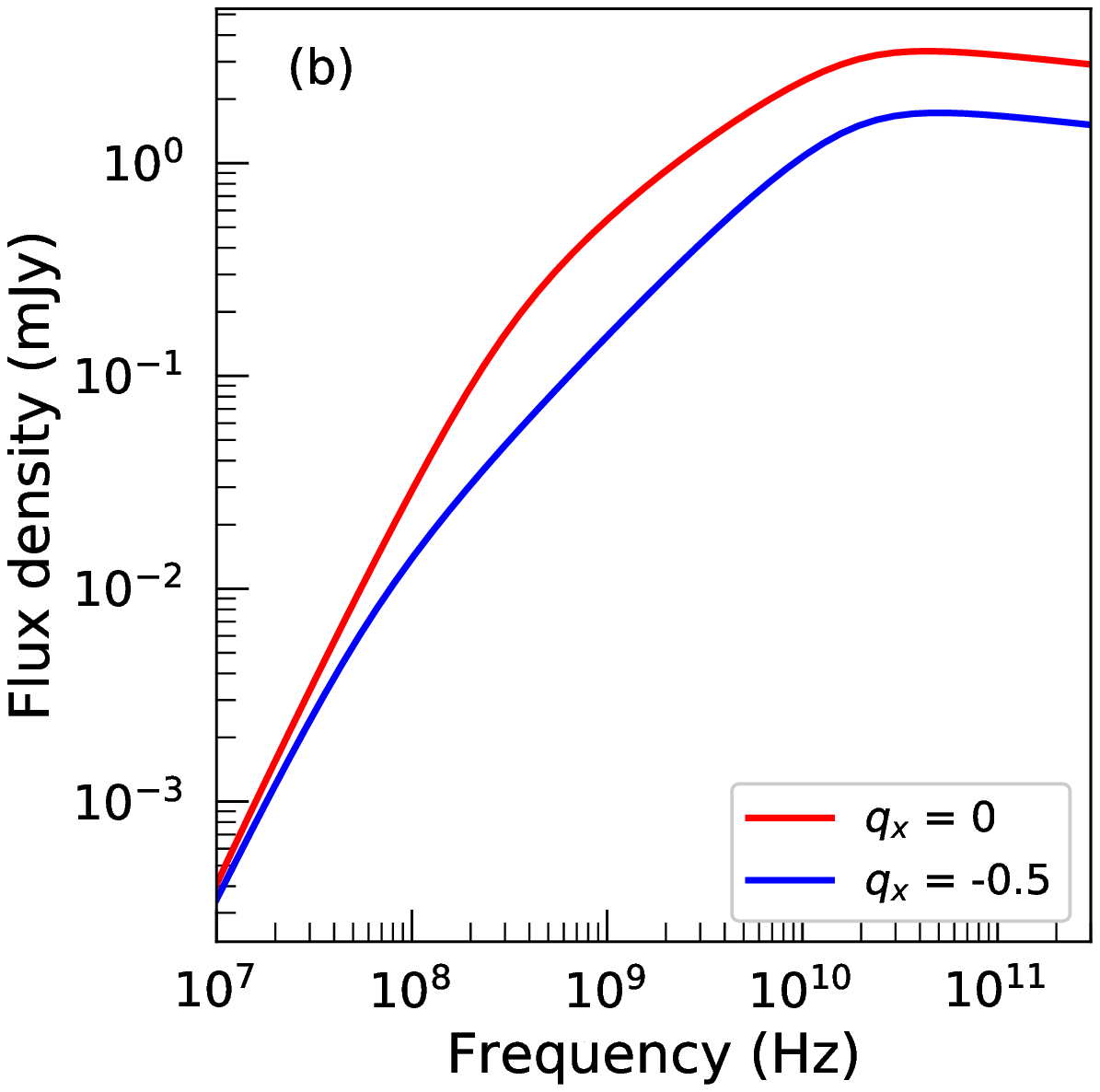}
        \hspace*{-0.8cm} 
        \includegraphics[width = 0.55\textwidth]{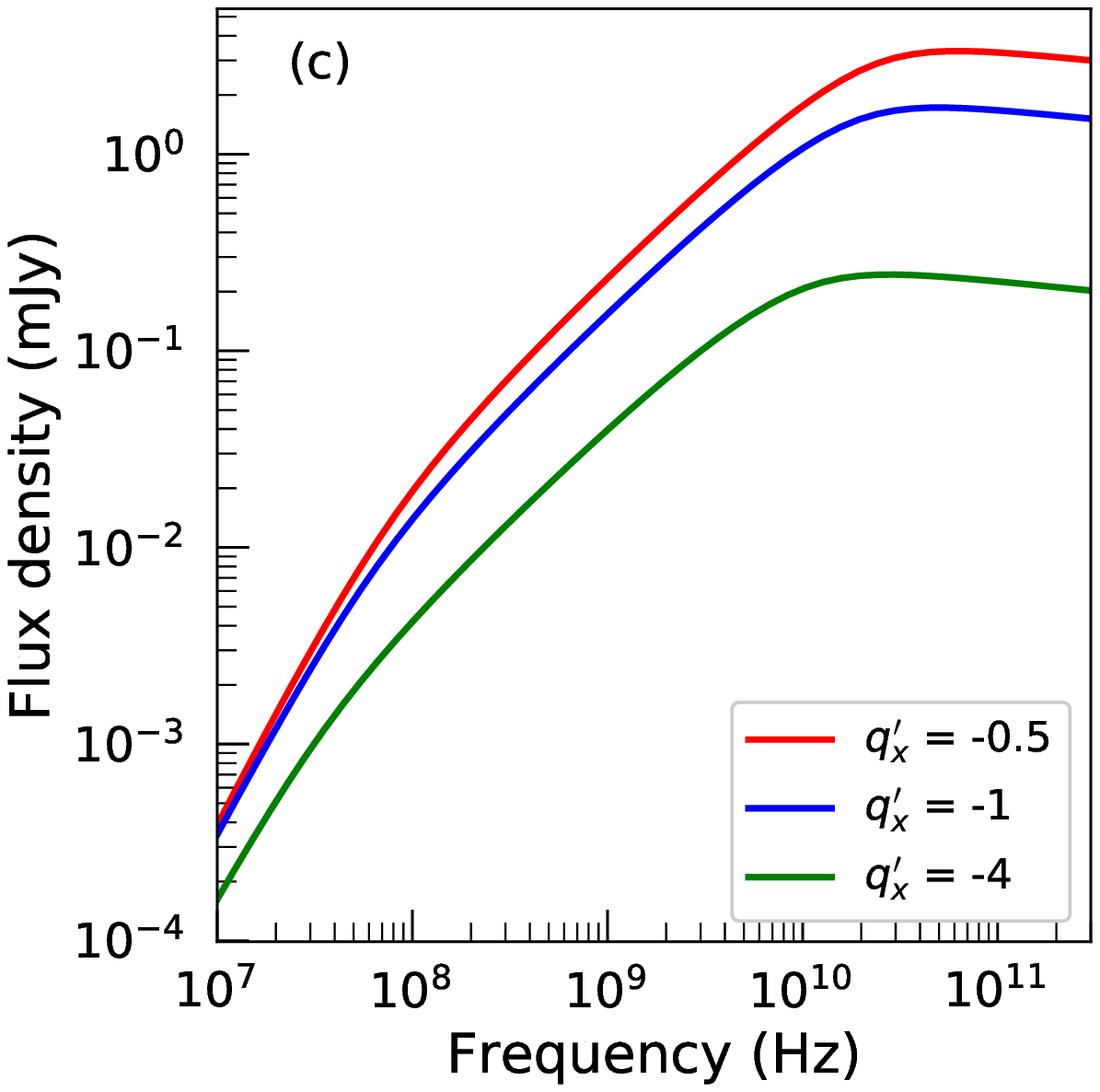} 
        \includegraphics[width = 0.55\textwidth]{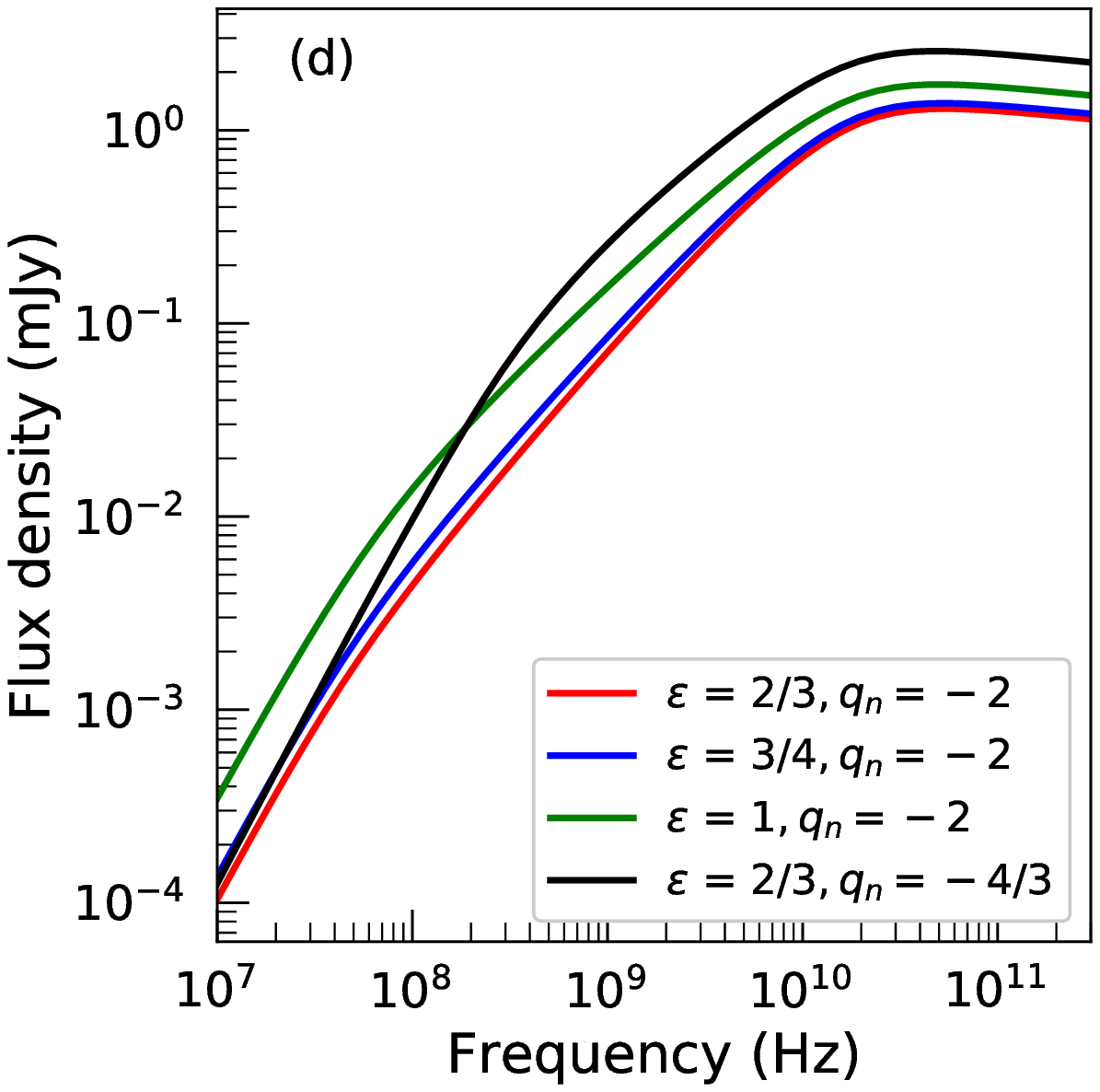}
        \begin{center}        
        \caption{Dependence of the radio spectrum of a fully thermal jet on (a) $x_0$, (b) $q_x$, (c) $q'_x$, (d) combination of $\epsilon$ and $q_n$. The parameters other than the ones varied here are same as that of Fig.~\ref{fig:set1}. The details of the spectra shown in (a), (b), (c) and (d) are listed in Table~\ref{tab:freeparms1} as Cases~IV, V, VI and VII, respectively.}
        \label{fig:set2}
         \end{center}
    \end{minipage}\hfill    
\end{figure}


\begin{figure}
	\hskip -0.3cm
	\includegraphics[width=\columnwidth]{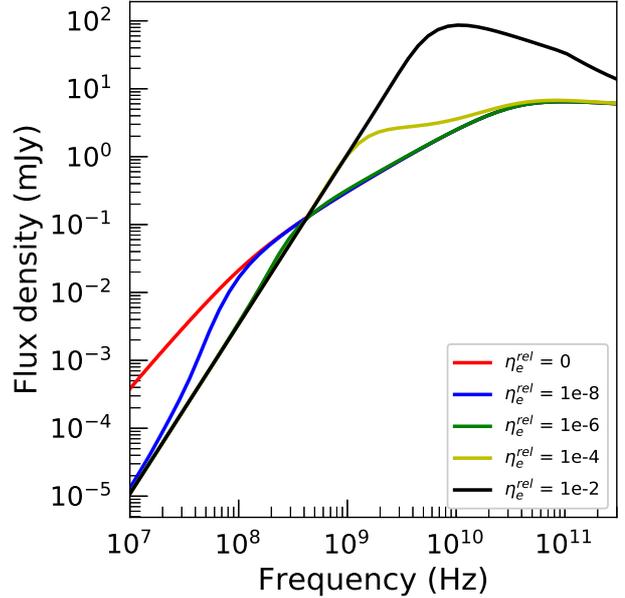}
	\vspace*{-2mm}
    \caption{Evolution of the radio spectrum of a knot with inner thermal region and outer non-thermal shell as a function of relativistic electron fraction ($\eta^{rel}_{e}$). The jet number density at $r_0$ is $n_0=10^8$~cm$^{-3}$, $\epsilon=1$, $q_n=-2$, and the remaining parameters of the thermal emission are same as those that remain fixed in the plots shown in Fig.~\ref{fig:set1}. The parameters of the non-thermal emission are $p$ = 2.3, $\delta \theta$ = 0.5$\degr$ and ${B_0} = 0.3$~mG. The details of the spectra are listed in Table~\ref{tab:freeparms1} as case~VIII.}
    \label{fig:set3}
\end{figure}


\par Fig.~\ref{fig:set2}~(a) displays the jet spectra for different ionization fractions $x_0$ at $r_0$~(Case~IV). The jet parameters are $n_0 = 5\times 10^7$~cm$^{-3}$ and the remaining parameters are same as that of Case~I. Since $x_0$ determines the ionized fraction of particles in the jet material that contribute to the emission, the flux densities increase with increase in $x_0$ as shown in the figure. 
The same reasoning is applicable to an increase in turnover frequencies with increase in $x_0$. We note that the Reynolds model predicts that the transition to a fully optically thin jet (high frequency turnover) is dependent on $x_0$ as $x_0^{2/2.1}$ suggesting an increase in turnover frequency with $x_0$. The spectral indices and turnover frequencies corresponding to the case explored here are listed as Case~IV in Table~\ref{tab:freeparms1}.

\par In Fig.~\ref{fig:set2}~(b), (c) and (d), we have shown the dependence of the spectrum on the power-law indices $q_x$, $q'_x$ and a combination of $\epsilon$ and $q_n$~(Cases~V, VI and VII, respectively). The jet parameters are $n_0 = 5\times 10^7$~cm$^{-3}$ and the remaining parameters other than the ones varied here are same as Case~I. From the plots, it is evident that flux densities decrease with increase in the value of $|q_x|$ and $|q'_x|$ as this results in the lowering of the number density of particles that are contributing to the emission. The opposite is true for $\epsilon$. To understand this better, we first consider the case of varying $q_x$. For this, we choose an arbitrary frequency of 1.5~GHz where our model is comparable with that of Reynolds. At this frequency, the ratio of flux densities for the Reynolds model between $q_x= 0$ and $q_x= -0.5$ is 3.6, whereas for our model the corresponding ratio is 3.3. This shows the agreement with the model of Reynolds at low frequency. 
The influence of $q'_x$ on the spectrum, shown in Fig.~\ref{fig:set2}(c), is expected to be such that the regions of the jet farther away from the axis will have lower number density of particles contributing to the emission, thereby reducing the overall flux density. The spectra agree with this prediction. The next power-law index that we are interested in studying is $\epsilon$ which plays a role in deciding the jet geometry. In Fig.~\ref{fig:set2}~(d), the red, blue and green curves represent a case of constant $q_n$ but varying $\epsilon$. The observed trend implies that the flux densities increase with an increase in the value of $\epsilon$ due to the jet becoming wider, and therefore assimilating more material that can contribute to the overall emission. At a frequency of 500~MHz, the ratio of flux densities for $\epsilon$ = 1 and 3/4 with respect to $\epsilon$ = 2/3 are 2.5 and 1.2, respectively. The corresponding values for Reynolds model are 2.9 and 1.3, respectively. 

Additionally, we have included a scenario in which both $\epsilon$ and $q_n$ are varied (black curve) in Fig.~\ref{fig:set2}~(d). A comparison with the other combinations in the figure shows that the low frequency turnover in this case occurs at a higher frequency while the high frequency turnover is similar to the other cases. From the red, blue and green curves in this figure, we note that increasing $\epsilon$ leads to an increase in flux densities, while for $|q_n|$ the effect is opposite (see Case~II). In Case~II, the flux densities at low frequencies where the full jet is optically thick do not vary significantly with a change in $q_n$, but vary considerably at high frequencies. Varying $\epsilon$ and $q_n$ simultaneously (black curve versus the others), the effect of varying $ q_n$ is minimal and the $\epsilon$ factor appears to dominate at low frequencies as a wider jet contributes to larger flux densities. At higher frequencies, a narrower jet with a flatter $q_n$ profile contributes to higher flux densities than a wider jet with similar or larger value of $|q_n|$ suggesting that $q_n$ appears to be the dominant parameter compared to $\epsilon$ here.  
The spectral indices and the associated turnover frequencies are listed in Table.~\ref{tab:freeparms1} as Case~V, VI and VII are listed in Table~\ref{tab:freeparms1}.

%


\begin{figure}
\hspace*{-0.2cm}
    \begin{minipage}{0.5\textwidth}
        \hspace*{-1cm}     
        \includegraphics[width = 0.55\textwidth]{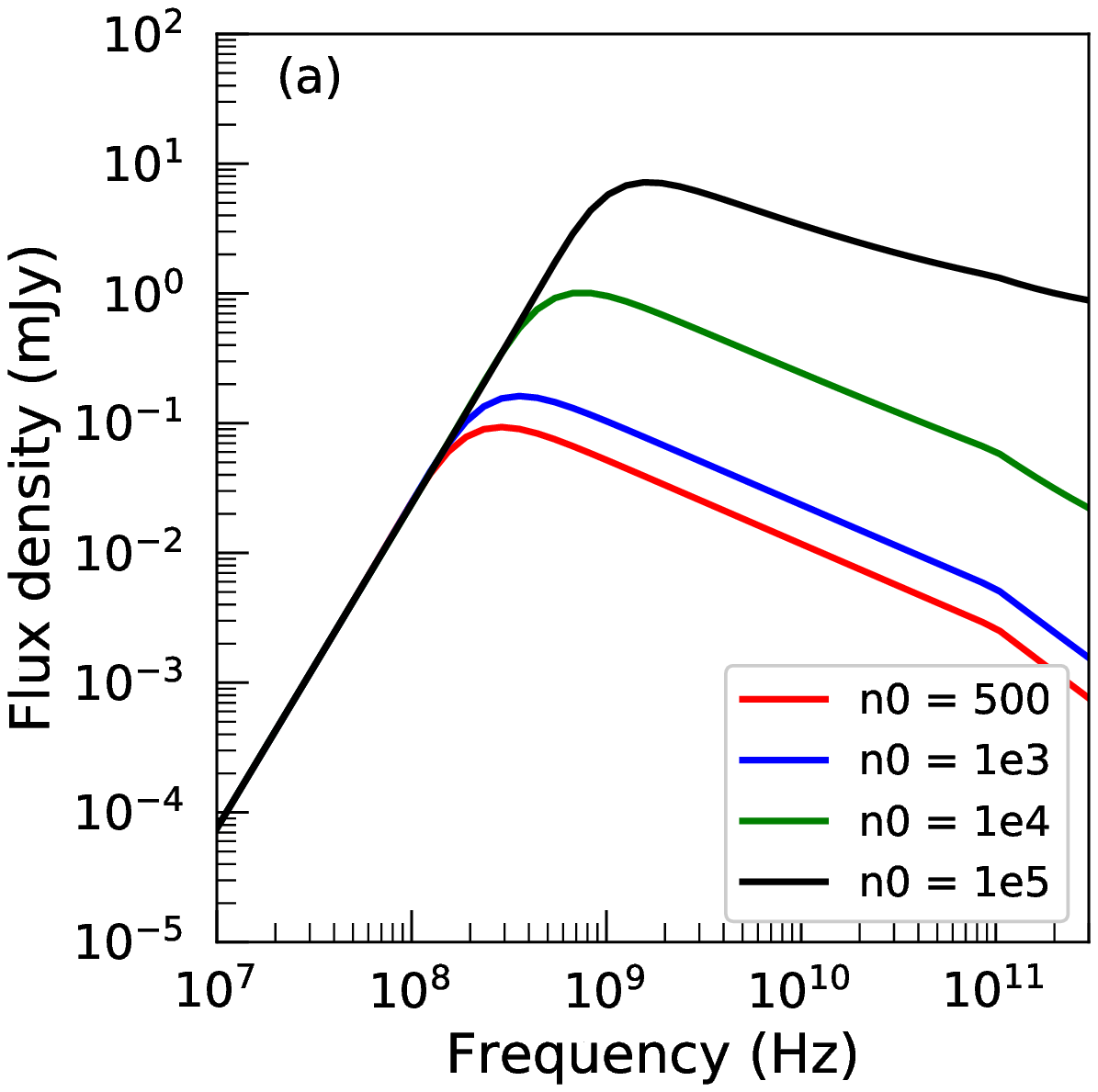}        
        \includegraphics[width = 0.54\textwidth]{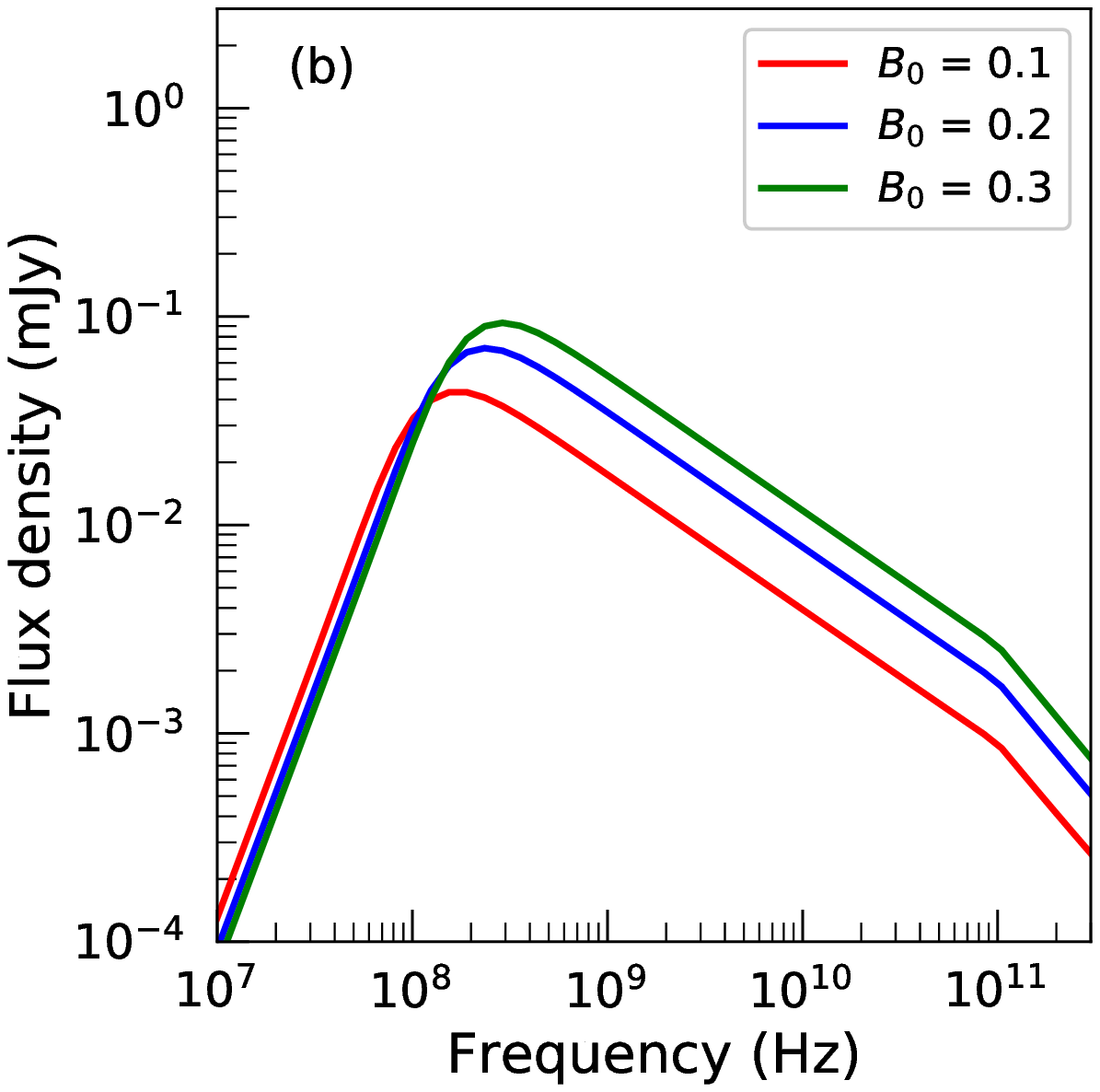}
        \hspace*{-1cm}
        \includegraphics[width = 0.55\textwidth]{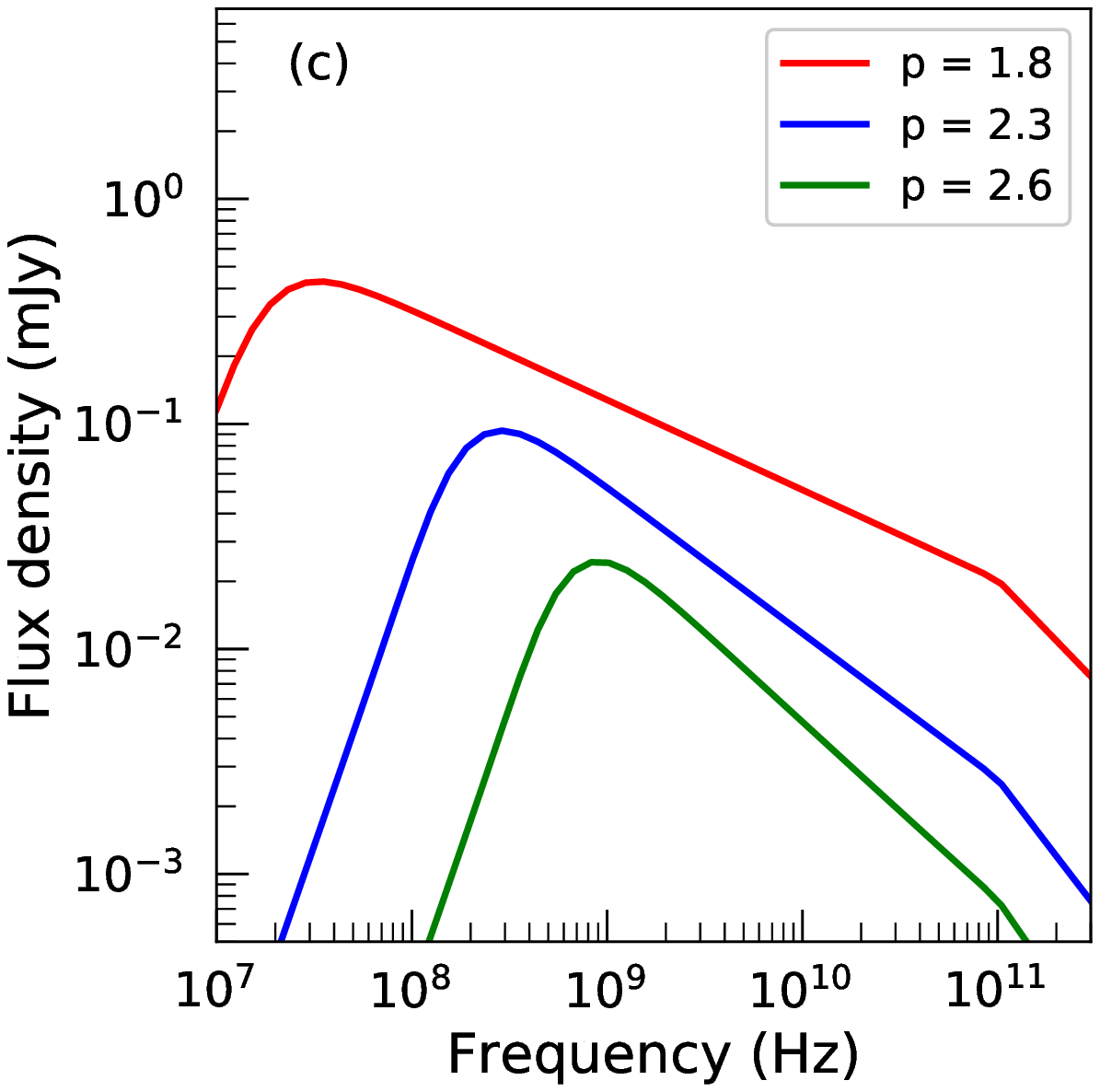} 
        \includegraphics[width = 0.55\textwidth]{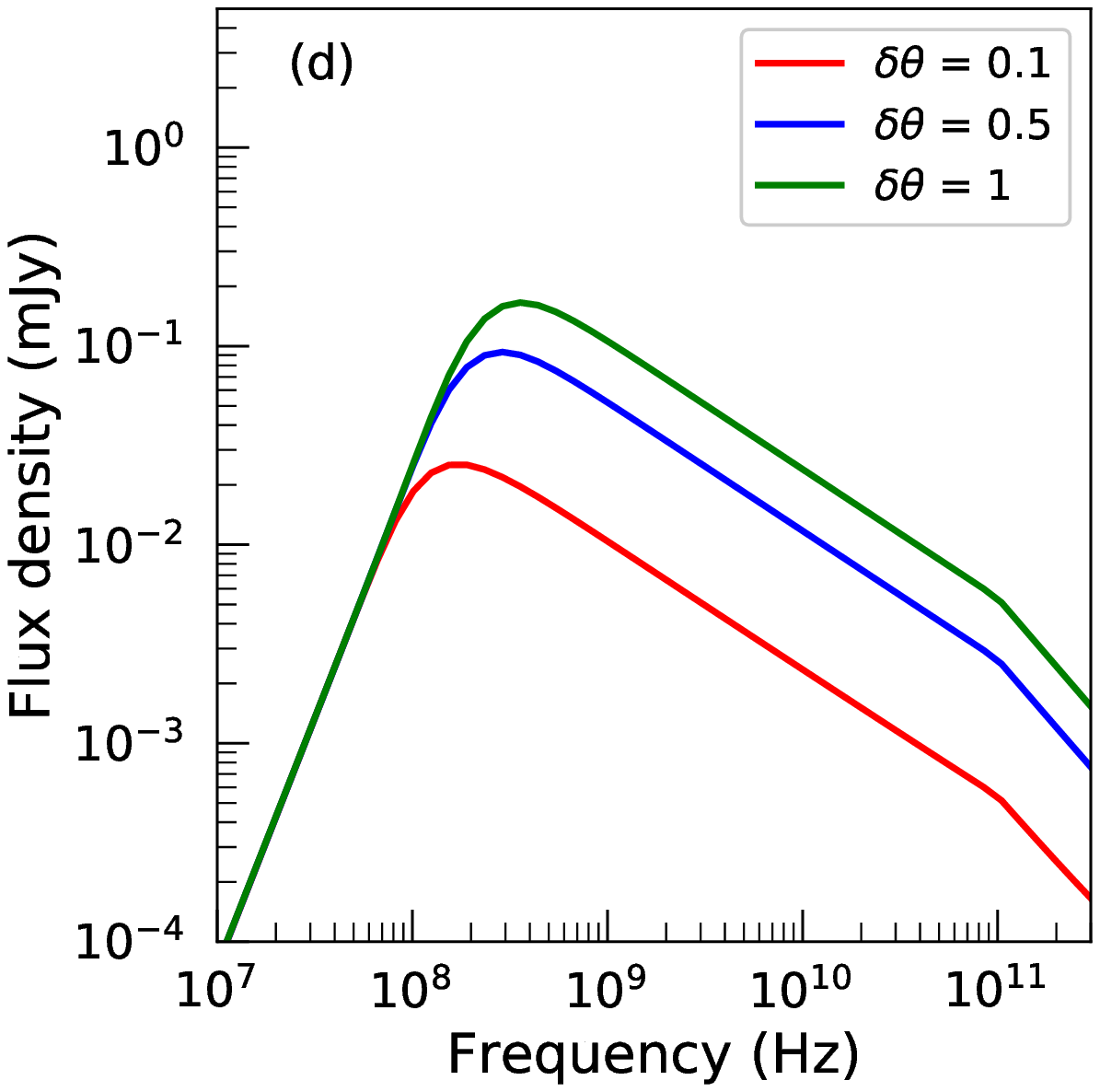}
        \begin{center}        
        \caption{Dependence of the radio spectrum of a knot with inner thermal region and an outer non-thermal shell on (a) $n_0$ $(\text{cm}^{-3}) $, (b) $B_0$~(mG), (c) $p$, and (d) $\delta \theta (\degr)$. We have used $r_0= 3000$~au and $y_{max} \sim 4500$~au. The remaining parameters other than the ones varied here are same as that of Fig.~\ref{fig:set1}. The parameters of synchrotron emission are $\delta \theta = 0.5\degr$, $p = 2.3$ , $B_0 = 0.3$~mG and $\eta^{rel}_{e} = 10^{-5}$. The details of the spectra shown in (a), (b), (c) and (d) are listed in Table~\ref{tab:freeparms1} as Cases~IX, X, XI and XII, respectively.}
         \label{fig:set4}
         \end{center}
    \end{minipage}\hfill   
\end{figure}



\begin{table*}
\caption{Turnover frequencies and associated spectral indices for fully thermal jet close to the YSO, and knots emitting combined thermal and non-thermal emission located farther away from the exciting YSO. The corresponding spectra are plotted in Figs~\ref{fig:set1},~\ref{fig:set2},~\ref{fig:set3} and \ref{fig:set4}.}
\begin{center}
\begin{threeparttable}

\begin{tabular}{ ccccc}
\hline \hline
  \multicolumn{5}{c}{\multirow{2}{*}{THERMAL FREE-FREE EMISSION}} \\
       &  & & \\
 \hline \hline
Variable Parameter & Value & &Turnover frequencies (GHz) & Spectral indices \\
\hline
\hline
  \multicolumn{5}{c}{ Case I [Fig.~\ref{fig:set1}(a)] } \\\hline \hline
 \multirow{2}{*}{ $n_0$ } & $10^6$ & & 0.76 & 0.80; -0.09 \\
 &$10^7$ & & 0.04; 6.87 & 1.30; 0.83; -0.09 \\
 $(\text{cm}^{-3}) $ &$5 \times 10^7$ & & 0.18; 31.17 & 1.57; 0.83; -0.06 \\
 &$10^8$  & & 0.35; 60.59 & 1.65; 0.84; -0.04 \\
\hline \hline
 \multicolumn{5}{c}{ Case II [Fig.~\ref{fig:set1}(b)] } \\\hline \hline
 \multirow{3}{*}{ $q_n$} & -1 & & 1.67; 32.70 & 1.86; 0.36; -0.08 \\
 & -2 & & 0.18; 31.17 & 1.57; 0.83; -0.06 \\
 & -2.5 &  & 0.03; 30.58 & 1.43; 1.04; -0.06 \\ 
 \hline
\hline
\multicolumn{5}{c}{ Case III [Fig.~\ref{fig:set1}(c)]}  \\\hline \hline
 \multirow{2}{*}{$r_0$} & 10 & & 14.92 & 0.88; -0.07\\
   & 50 & & 0.18; 31.17 & 1.57; 0.83; -0.06\\
 (au) & 100 & & 0.86; 45.52 & 1.72; 0.76; -0.06\\
 & 500 & & 101.35 &  1.71; -0.01\\
 \hline
 \hline
  \multicolumn{5}{c}{ Case IV [Fig.~\ref{fig:set2}(a)] } \\\hline \hline
\multirow{3}{*}{$x_0$} & 0.02 & & 3.61 & 0.89; -0.08\\
& 0.2 & & 0.18; 31.17 & 1.57; 0.83; -0.06 \\
& 1 & & 0.84; 146.27 & 1.71; 0.83; -0.01\\
\hline  \hline
\multicolumn{5}{c}{ Case V [Fig.~\ref{fig:set2}(b)]}  \\\hline \hline
\multirow{2}{*}{$q_x$} & 0 &  &  1.81; 31.79 & 1.57; 0.49; -0.07\\
& -0.5 & & 0.18; 31.17 & 1.57; 0.83; -0.06\\
\hline \hline
 \multicolumn{5}{c}{ Case VI [Fig.~\ref{fig:set2}(c)]}  \\\hline \hline
\multirow{3}{*}{$q'_{x}$}& -0.5 & & 0.23; 38.18 & 1.64; 0.85; -0.06 \\
& -1 & & 0.18; 31.17 & 1.57; 0.83; -0.06 \\
& -4 & & 0.09; 18.73 & 1.44; 0.82; -0.07 \\
\hline
\hline
 \multicolumn{5}{c}{ Case VII [Fig.~\ref{fig:set2}(d)]}  \\\hline \hline
 \multirow{3}{*}{$\epsilon$, $q_n$} & 2/3, -2 & & 0.07; 30.87 & 1.70; 1.05; -0.06 \\
 & 3/4, -2 & & 0.08; 30.87 & 1.68; 0.99; -0.06 \\
 & 1, -2 & & 0.18; 31.17 & 1.57; 0.83; -0.06\\
  & 2/3, -4/3 & & 0.69; 31.77 & 1.78; 0.71; -0.07\\
\hline  \hline

 \multicolumn{5}{c}{\multirow{2}{*}{THERMAL FREE-FREE AND NON-THERMAL SYNCHROTRON EMISSION}} \\
       &  & & \\
 \hline \hline
 \multicolumn{5}{c}{ Case VIII [Fig.~\ref{fig:set3}] } \\\hline \hline
\multirow{5}{*}{$\eta^{rel}_{e}$} & 0 & & 0.10; 95.46 & 1.82; 0.87; -0.07  \\
&$10^{-8}$ & & 0.16; 95.28 & 3.14; 0.83; -0.07  \\
&$10^{-6}$ & & 0.36; 95.15  & 2.55; 0.78; -0.07  \\
&$10^{-4}$ & & 1.92; 85.02  &  2.44; 0.36; -0.09 \\
&$10^{-2}$ & & 10.69; 100 & 2.40; -0.43; -0.76  \\
\hline \hline
  \multicolumn{5}{c}{ Case IX [Fig.~\ref{fig:set4}(a)] } \\\hline \hline
\multirow{2}{*}{$n_0$} &$500$ & & 0.29; 100 & 2.31; -0.63; -1.13 \\
 &$10^3$ & & 0.36; 100 & 2.26; -0.64; -1.12 \\
$(\text{cm}^{-3}) $ &$10^4$ & & 0.75; 100 & 2.36; -0.61; -0.89 \\
 &$10^5$ & & 1.63; 100 & 2.39; -0.43; -0.35 \\ 
\hline
\hline
  \multicolumn{5}{c}{ Case X [Fig.~\ref{fig:set4}(b)] } \\\hline \hline
\multirow{2}{*}{$B_0$ } & 0.1 & & 0.17; 100 &  2.20; -0.64; -1.10  \\
& 0.2 & & 0.24; 100 &  2.20; -0.64; -1.1 \\
(mG) & 0.3 & & 0.29; 100 & 2.31; -0.63; -1.13  \\
\hline \hline
 \multicolumn{5}{c}{ Case XI [Fig.~\ref{fig:set4}(c)] } \\\hline \hline
\multirow{3}{*}{$p$} & 1.8 & & 0.03; 100 &  1.18; -0.40; -0.90 \\
& 2.3 & & 0.29; 100 & 2.31; -0.63; -1.13 \\
& 2.6 & & 0.90; 100 &  2.39; -0.77; -1.23  \\
\hline \hline
 \multicolumn{5}{c}{ Case XII [Fig.~\ref{fig:set4}(d)] } \\\hline \hline
\multirow{2}{*}{$\delta \theta$} & 0.1 & & 0.17; 100 &  2.19; -0.63; -1.06 \\
& 0.5 & & 0.29; 100 & 2.31; -0.63; -1.13 \\
$(\degr)$ & 1 & & 0.37; 100 &  2.26; -0.64; -1.14 \\
\hline
\end{tabular}
\label{tab:freeparms1}

\end{threeparttable}

\end{center}
\end{table*}

\subsection{Jet with combination of thermal and non-thermal emission}\label{sub:th_nt}

\par As we move farther away from the central YSO, the densities are expected to decrease thereby reducing the thermal contribution in the jet spectrum, and resulting in the increased contribution of non-thermal emission.  To describe this evolution, we have considered a jet with parameters $r_0 = 50$~au, $y_{max} = 1300$~au, $n_0 = 10^8$~cm$^{-3}$, $q_n= -2$, $\theta_0 = 30^\circ$, $\epsilon= 1$, $i = 60^\circ$, $x_0= 0.2$, $q_x= -0.5$, $q'_x= -1$, $T_0 = 10^{4}$~K at a distance $d =~1$~kpc. We have included the effects of synchrotron emission in the outer edges of the jet with an angular thickness of $\delta \theta = 0.5\degr$. The values of parameters related to the synchrotron emission are $p= 2.3$ and $B_0 = 0.3$~mG. The evolution of jet spectrum from dominating thermal emission to dominating synchrotron emission is illustrated by varying the parameter $\eta^{rel}_{e}$ (Case~VIII). It can be seen from Fig.~\ref{fig:set3} that as $\eta^{rel}_{e}$ increases, non-thermal emission gradually dominates over thermal emission. During the course of this evolution, we observe a few turnover frequencies relating the optically thick/thin emission from the thermal/synchrotron mechanisms, as described in Sect.\ref{sync}. The spectral indices and turnover frequencies for this case are listed in Table~\ref{tab:freeparms1} as Case~VIII.

\par We next consider the case of a knot located farther away from the exciting YSO to elucidate the effect of the parameters of synchrotron mechanism on the jet spectrum. In this case, we take parameters characteristic of a farther knot with $r_0= 3000$~au, $y_{max}\sim4500$~au, $n_0=500$~cm$^{-3}$, $q_n$ = -2, $\theta_0 = 30^\circ$, $\epsilon = 1$, $i = 60^\circ$, $x_0 = 0.2$, $q_x = -0.5$, $q'_x = -1$, $T_0 = 10^{4}$~K at a distance $d =~1$~kpc. We take $\delta \theta = 0.5\degr$, $p = 2.3$, $B_0 = 0.3$~mG and $\eta^{rel}_{e} = 10^{-5}$. We first study the effect of number density on the jet spectrum. This is shown in Fig.~\ref{fig:set4}~(a)~(Case~IX). Similar to what was observed in the thermal jet case, with an increase in $n_0$ at $r_0$, we find that the flux densities increase, due to increase in the number density of particles contributing to the overall emission. We also note that with a decrease in $n_0$ (i) the low frequency turnovers shift to lower values due to the jet becoming more optically thin overall, and (ii) the optically thin spectral index becomes steeper due to the gradual domination of synchrotron emission over thermal emission. Note that the variation of flux density with number density is not the same in the two cases; it varies as $n_0^2$ and $n_0$ for thermal and non-thermal emission, respectively. The details of the spectra are listed in Table~\ref{tab:freeparms1}. The other panels of the figure encapsulate the effect of the synchrotron parameters $B_0$~(Case~X), $p$ (Case~XI) and $\delta \theta$ (Case~XII) on the spectrum of the knot. From the figure it is observed that, the flux densities increase with increase in the values of $B_0$ and $\delta \theta$. The strength of $B_0$ determines the degree of acceleration imparted to the electrons through the Lorentz force. This in turn dictates the amount of energy delivered to them which will eventually be emitted in the form of synchrotron radiation. Likewise, an increase in $\delta \theta$ implies an increase in the angular thickness of the region from which synchrotron emission arises, thereby contributing to higher synchrotron flux densities. On the contrary, an increase in $p$ leads to a reduction in flux densities, and the optically thin spectral index is strongly dependent on the value of $p$. This is because as $p$ increases, the number of electrons in the higher energy regime decreases, thereby lowering the corresponding emission by these electrons. The plots shown in the figure support the above arguments. The spectral index values and turnover frequencies are listed in Table~\ref{tab:freeparms1} as Case~IX, X, XI and XII.


\begin{figure}
 
    \begin{minipage}{0.5\textwidth}
        \hspace*{-0.7cm}  
        \includegraphics[width = 0.55\textwidth]{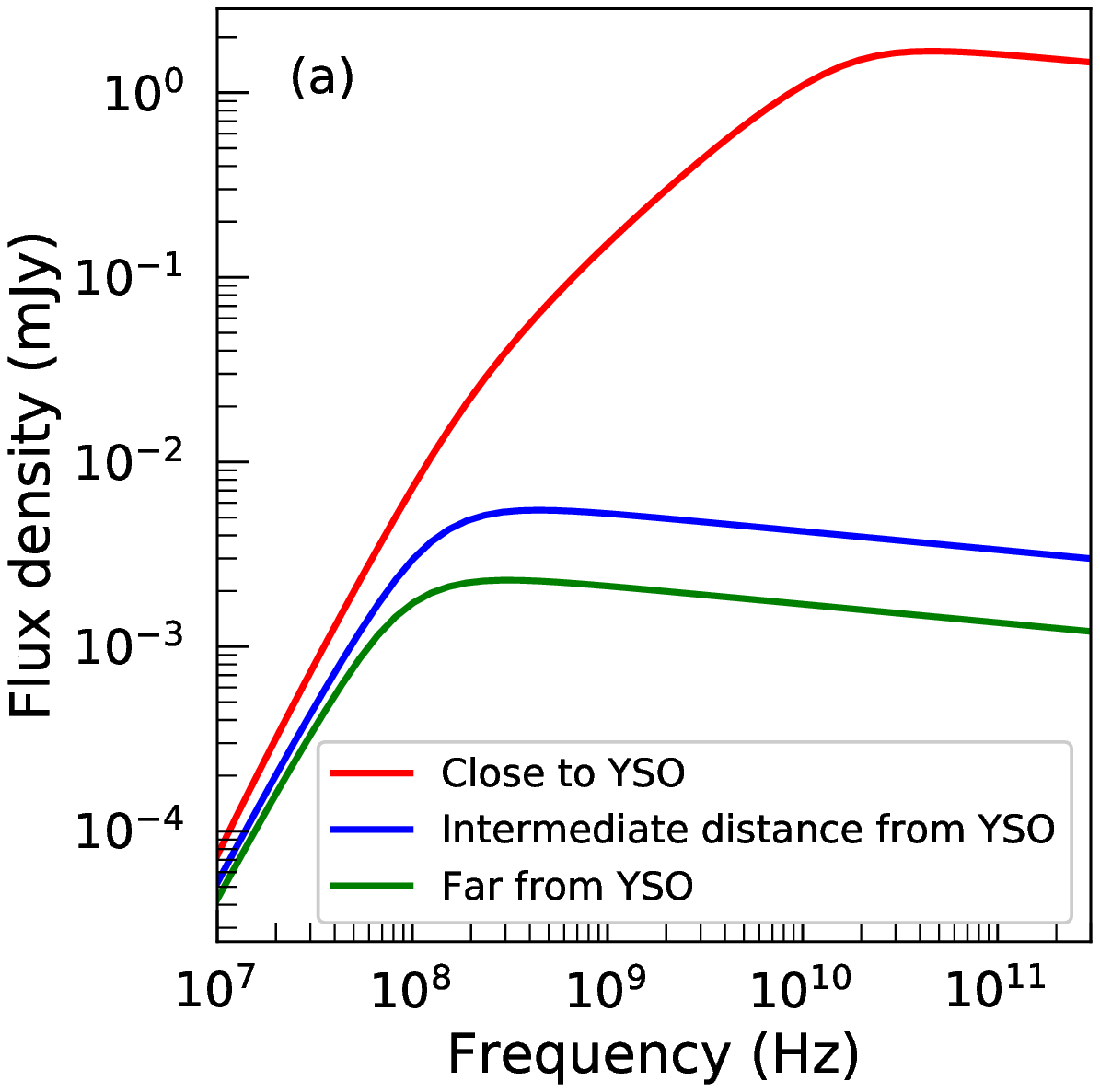}       
        \includegraphics[width = 0.55\textwidth]{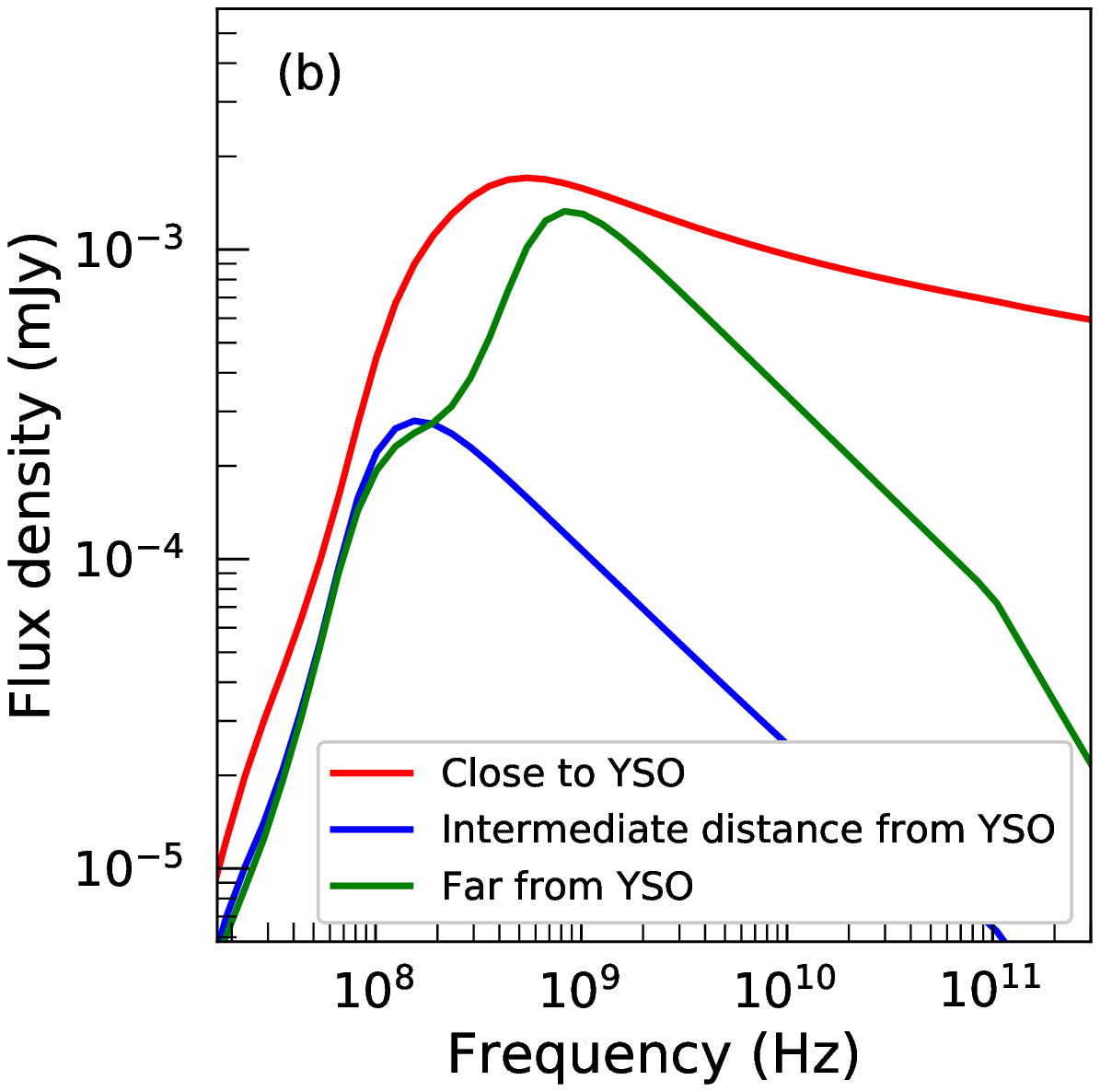}
        \begin{center}        
        \caption{Radio spectrum of (a) a fully thermal jet, and (b) a jet with with inner thermal region and an outer non-thermal shell, showing emissions from three different regions of equal volume along the jet length. The inclination angle of the jets are $i = 90^\circ$. The remaining parameters of the jets in (a) and (b) are listed in the text where this is discussed. The details of the spectra shown in (a) and (b) are listed in Table~\ref{tab:freeparms2} as Cases~XIII and XIV, respectively.} 
         \label{fig:set5}
         \end{center}
    \end{minipage}\hfill   
\end{figure}



\hspace*{-2cm}
\begin{table*}
\caption{Turnover frequencies and associated spectral indices for fully thermal jet close to the YSO, and a jet emitting combined thermal and non-thermal emission located farther away from the exciting YSO, i.e. Cases~XIII and XIV. The corresponding spectra are plotted in Fig.~\ref{fig:set5}.}
\begin{center}
\begin{threeparttable}
\begin{tabular}{ lccc}
\hline \hline
    \multicolumn{4}{c}{\multirow{2}{*}{THERMAL FREE-FREE EMISSION}} \\
       &  & & \\
 \hline \hline
   \multicolumn{4}{c}{ Case XIII [Fig.~\ref{fig:set5}(a)] } \\\hline \hline
Jet region & & Turnover frequencies (GHz) & Spectral indices \\
\hline
\hline
Close to YSO & & 0.23; 46.03 & 1.91; 0.81; -0.08 \\
Intermediate distance from YSO & & 0.44 & 1.29; -0.09 \\
Farther from YSO & & 0.31 & 1.24; -0.09 \\
\hline  \hline
 \multicolumn{4}{c}{\multirow{2}{*}{THERMAL FREE-FREE AND NON-THERMAL SYNCHROTRON EMISSION}} \\
       &  & & \\

 \hline \hline
    \multicolumn{4}{c}{Case XIV [Fig.~\ref{fig:set5}(b)] } \\\hline \hline
Close to YSO & & 0.48 & 1.81; -0.17 \\
Intermediate distance from YSO & & 0.20; 100.00 & 1.90; -0.62; -0.83 \\
Farther from YSO  & & 0.19; 0.83; 100.00 & 2.01; $-^*$; -0.64; -1.14 \\
 \hline
 \hline
 
\end{tabular}
\label{tab:freeparms2}
\begin{tablenotes}
\scriptsize{$*$ The spectral index is not well-defined in this region due to the close proximity of two turnover frequencies.}
\end{tablenotes}
\end{threeparttable}

\end{center}
\end{table*}


\subsection{Emission across the jet length}\label{sub:emission_nature_l}

\par To understand the relative contributions of different regions within the jet to the overall spectrum, we divide the jet into three regions of equal volume along its length. For simplicity, we assume the jet to be in the plane of the sky i.e with an inclination angle $i = 90^\circ$. The first region which we consider is close to the injection radius where the number densities and optical depths are very high. We then take the intermediate and farther regions along the jet length where the number densities gradually decrease according to the radial power-law index $q_n$. Fig.~\ref{fig:set5}~(a) displays the spectra of the three regions in the case of a fully thermal jet with parameters $r_0 = 50$~au, $\theta_0 = 30^\circ$, $n_0 = 5\times10^7$~cm$^{-3}$, $q_n = -2$, $x_0 = 0.2$, $q_x = -0.5$, $q'_x = -1$, $T_0 = 10^4$~K, $q_T = 0$, $\epsilon = 1$ and $y_{max} = 700$~au at a distance d = 1~kpc (Case~XIII). For this fully thermal jet, the region close to the YSO being highly optically thick, displays two turnover frequencies in the frequency window of interest. On the other hand, as we move farther away, due to decrease in number density and optical depth, both the turnover frequencies shift to lower values. In both the cases, the low frequency turnover goes below the frequency range displayed and the total flux contributions also reduce. The spectral indices and turnover frequencies associated with this case are listed in Table~\ref{tab:freeparms2} as Case~XIII.

\par We next consider a jet that includes a combination of thermal and non-thermal emission and Fig.~\ref{fig:set5}~(b) shows the spectra of the three regions of the jet (Case~XIV). In this case, we assume a jet located at $r_0 = 100$~au and $y_{max} = 450$~au with a density $n_0 = 10^6$~cm$^{-3}$ and the remaining parameters are the same as that of Case~XIII. This jet includes shocked edges of thickness $\delta \theta = 0.5^\circ$, where the parameters for synchrotron emission are $B_0 = 0.3$~mG, $\eta_{rel} = 10^{-5}$ and $p = 2.3$. It is evident from the figure that the region close to the YSO is dominated by thermal free-free emission, the intermediate region shows competing contributions from both thermal and non-thermal mechanisms while the farthest region in the jet shows dominant non-thermal emission. As we move from the region close to the YSO to the intermediate region, the number density declines and this manifests as a decrease in flux densities and turnover frequencies. This can be seen from the figure. In both these regions, the contribution to synchrotron emission arises from the shocked regions at lateral edges of the jet. On the other hand, in the region of the jet farthest from the central YSO, apart from the lateral edges, there is an additional contribution of synchrotron emission from the top terminal edge. As a result, the flux densities and turnover frequencies increase compared to the intermediate region of the jet. The figure emulates this behaviour. The spectral indices and turnover frequencies associated with this case are listed in Table~\ref{tab:freeparms2} as Case~XIV.  \\

Thus, we have explored the effect of various model parameters on the radio spectra of jets. This is aimed at achieving a better understanding of the behavior of the jet model in different plausible scenarios. Such a characterization would enable the application of the model to observations of various jets to derive fundamental parameters that are unlikely to be obtained directly from observations. In this context, the two-fold application of the model as mentioned in Sect.~\ref{model} will aid the study jets/knots which are both close and far away from the exciting YSO.

\section{Summary} \label{summary}
Protostellar jets are ejected during the accretion phase of the formation of a star, and are highly collimated. Observations indicate that jets could manifest in the form of knots. In general, the observed radio spectrum of these knots can be explained as a combination of thermal (free-free) and non-thermal (synchrotron) emission processes. The most widely employed model for thermal jets is the Reynolds model which utilizes the small opening angle approximation and calculates the spectral index of a standard conical jet for which a part of the jet is optically thick and part is optically thin, to be +0.6. This model also computes other cases of jet spectral indices with power-law profiles for density, temperature, ionization fraction and opening angle. In our model, we have introduced a wide-angle geometry and a generalized optical depth compared to the Reynolds model. There are a few additional features, the most important being the synchrotron generation in the shocks formed at the jet lateral edges where the relativistic population of electrons contributes to synchrotron and the non-relativistic population contributes to free-free emission. Along with the radial decrease in ionization fraction, we have also introduced a decrease of ionization fraction laterally across the jet cross-section. We have explored the change in turnover frequencies and spectral indices associated with them for variations in different physical parameters of the radio jet.

\section{Acknowledgemnts}
We thank the referee for the valuable comments and suggestions that have helped us to improve the quality of this work. 

\section*{Data Availability}
The numerically generated data and codes underlying this article will be shared on reasonable request to the corresponding author.

\bibliography{radiopjet_ref}

\appendix

\section{JET GEOMETRY FOR VARYING OPENING ANGLE}\label{ap:varyingth}

\begin{figure}
	\hskip -0.4cm
	\includegraphics[width=\columnwidth]{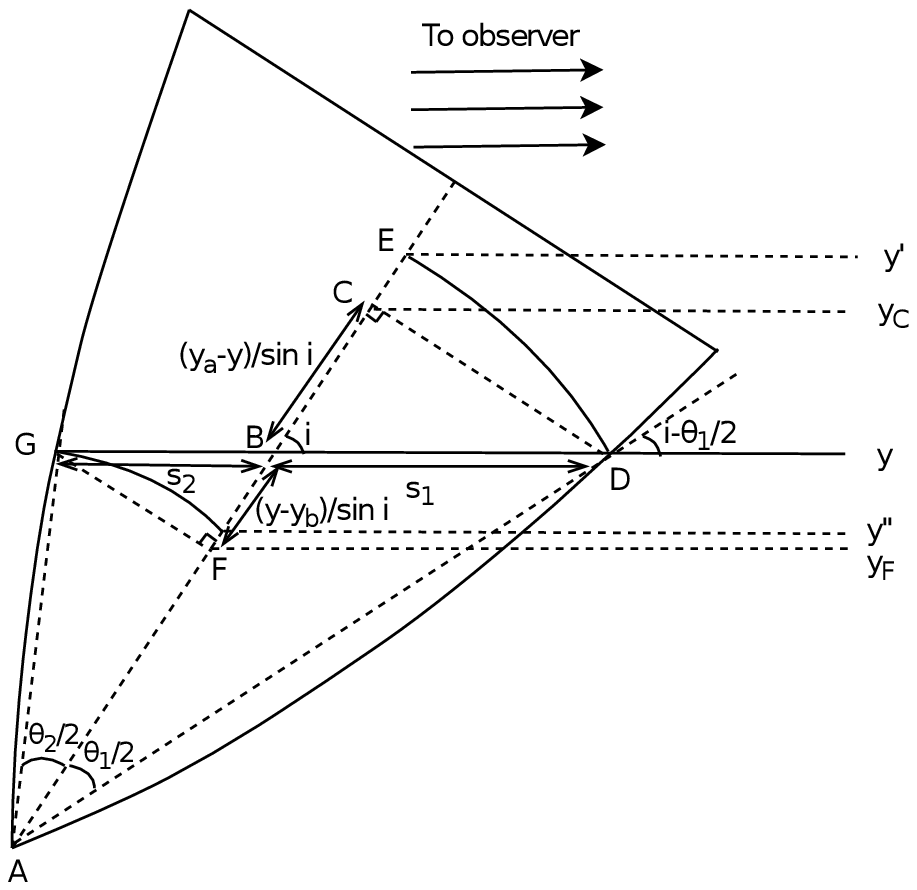}
    \caption{Schematic diagram of a jet for which the opening angle decreases with y ($\epsilon < 1$). Shown here is a LOS at y passing through the main axis of the jet. Here $\frac{\theta_1}{2}$ and $\frac{\theta_2}{2}$ are the opening angles corresponding to the jet front and rear edges, respectively.}
    \label{fig:vary_th}
\end{figure}

Fig.~\ref{fig:vary_th} shows the schematic diagram of a jet with varying opening angle for collimated case of $\epsilon < 1$, and represented here is the LOS passing through the long axis of the jet at an arbitrary projected length $y$. Owing to the inclination with respect to the plane of the sky, for any jet with $\epsilon \neq 1$, the opening angle corresponding to the front and rear edges of the jet will be different. As evident from the figure, the opening angles where the LOS crosses the front and rear edges of the jet are $\frac{\theta_1}{2}$ and $\frac{\theta_2}{2}$, respectively. $\theta_1$ is smaller and $\theta_2$ is larger than $\theta$. In order to calculate the two opening angles we first look into the geometry of the jet. In the figure, the base of the jet is represented by A, while B and D represents the intersection of LOS with the long axis and front edge of the jet, respectively. 
C is the base of the perpendicular from D to the axis and the arc DE represents all the elements within the jet located at the same radial distance from the central source. For any $y$, we define $y_{C}$ and $y'$ as the projected length of AC and AE on the sky plane, respectively.  From Eqn.~\eqref{eq:op_ang}, we know that any arbitrary $y'$ can be written as,
\begin{equation}
y' = y \bigg[ \frac{\tan(\theta_1/2)}{\tan(\theta/2)} \bigg]^{1/\epsilon\,-\,1}
\end{equation}
\noindent Similar to $y_{C}$ and $y'$, $y_{F}$ and $y"$ are defined for the rear edge of the jet. From the triangles ABD and CBD, it is clear that the side BD (=$s_1$) can be evaluated separately and then equated to obtain $\theta_1$ as follows. 
\begin{equation}\label{eq:theta1}
\bigg[ \bigg(\frac{\tan(\theta_1/2)}{\tan(\theta/2)}\bigg)^{1/\epsilon\,-\,1} -1 \bigg] - \frac{\sin(\theta_1/2)}{\sin(i-\theta_1/2)}\cos(i) = 0
\end{equation} 
\noindent Similar calculations for the rear half of the jet using triangles ABG and FBG, and equating BG (=$s_2$) gives $\theta_2$ as follows.
\begin{equation}\label{eq:theta2}
\bigg[1 - \bigg(\frac{\tan(\theta_2/2)}{\tan(\theta/2)}\bigg)^{1/\epsilon\,-\,1} \bigg] - \frac{\sin(\theta_2/2)}{\sin(i-\theta_2/2)}\cos(i) = 0
\end{equation} 

\noindent With the knowledge of the opening angles, the inclined widths of the front and behind half of the jet, $s_1$ and $s_2$ can be calculated in the following way.
\begin{equation}\label{eq:s1_curved}
s_1 = \frac{y\,\sin(\theta_1/2)}{\sin(i- \theta_1/2)\,\sin(i)}
\end{equation}
\begin{equation}\label{eq:s2_curved}
s_{2} = \frac{y\,\sin(\theta_2/2)}{\sin(i+ \theta_2/2)\,\sin(i)}
\end{equation}


\section{CALCULATION OF SYNCHROTRON SPECTRUM}\label{ap:sync_spec}

The source function and optical depth at the synchrotron characteristic frequency $\nu_{pk}$ are,
\begin{center}
$I_{\nu_{pk}}$ = $\frac{j_{\nu_{pk}}}{\alpha_{\nu_{pk}}}$
\end{center}

\begin{center}
$\tau_{\nu_{pk}}$ = $\int_{0}^{\mathscr{S}} \alpha_{\nu_{pk}} ds$
\end{center}
where $\mathscr{S}$ represents the LOS through the jet cross-section. The flux density at any observing frequency $\nu$ is given by,

\vspace{0.4cm}
For $\nu\leq\nu_{c}$ :\\
\hspace*{1cm} $\tau_{\nu}$ = $\tau_{\nu_{pk}}$ $\bigg(\frac{\nu}{\nu_{pk}}\bigg)^{-(p+4)/2}$\\
\hspace*{1cm} $S_{\nu}$ = $\displaystyle\int_{y_{0}}^{y_{max}}\int_{0}^{w(y)}\frac{2\, dw\, dy}{d^{2}}$ $I_{\nu_{pk}}$ ($\frac{\nu}{\nu_{pk}})^{5/2}$ (1 - $e^{-\tau_{\nu}}$)\\

For $\nu>\nu_{c}$ :\\
\hspace*{1cm} $\tau_{\nu}$ = $\tau_{\nu_{pk}}$ $\bigg(\frac{\nu_{c}}{\nu_{pk}}\bigg)^{-(p+4)/2}$ $\bigg(\frac{\nu}{\nu_{c}}\bigg)^{-(p+5)/2}$\\
\hspace*{1cm} $S_{\nu}$ = $\displaystyle\int_{y_{0}}^{y_{max}}\int_{0}^{w(y)}\frac{2\, dw\, dy}{d^{2}}$ $I_{\nu_{pk}}$ ($\frac{\nu}{\nu_{pk}})^{5/2}$ (1 - $e^{-\tau_{\nu}}$)\\

\section{CALCULATION OF FLUX DENSITY FROM AN ARBITRARY LOS}\label{ap:shell}

For an elemental material located in the shocked region at a distance $s$ along any LOS having a number density $n$, ionization fraction $x$ and fraction of relativistic electrons $\eta^{rel}_{e}$, the number density of electrons capable of emitting free-free radiation are given by,
\begin{equation}\label{eq:shell_ff}
\begin{split}
n^{ff} & = n\,x\,(1-\eta^{rel}_{e})\\
& = n_{0}\,x_0 \bigg(\frac{y'}{y_{0}}\bigg)^{q_n}\, \left(\frac{y'}{y_0}\right)^{q_x} \left[\frac{w(y') - w'(y')}{w(y')}\right]^{q_x'}\,(1-\eta^{rel}_{e})
\end{split}
\end{equation}
On the other hand, the number density of electrons capable of emitting synchrotron are given by,
\begin{equation}\label{eq:shell_nt}
\begin{split}
n^{syn} & = n\,x\,\eta^{rel}_{e}\\
& = n_{0}\,x_0 \bigg(\frac{y'}{y_{0}}\bigg)^{q_n}\, \left(\frac{y'}{y_0}\right)^{q_x} \left[\frac{w(y') - w'(y')}{w(y')}\right]^{q_x'}\,\eta^{rel}_{e}
\end{split}
\end{equation}

Note that the number density of free-free emitters from the fully thermal jet would correspond to $\eta^{rel}_{e}\,=\,0$ in Eqn.~\eqref{eq:shell_ff}. From Eqns.~\eqref{ff_jnu},~\eqref{ff_alphanu},~\eqref{eq:emission} and \eqref{eq:absorption} it is clear that for the shocked region, $j^{ff+syn}_{\nu}= A_{\nu,1}(n^{ff})^{2} + A_{\nu,2}{n^{syn}}$ and $\alpha^{ff+syn}_{\nu}= B_{\nu,1}(n^{ff})^{2} + B_{\nu,2}{n^{syn}}$, where $A_{\nu,1}$, $B_{\nu,1}$ and $A_{\nu,2}$, $B_{\nu,2}$ are constants associated with free-free and synchrotron mechanisms whereas for the case of a thermal jet both $j^{ff}_{\nu}$ and $\alpha^{ff}_{\nu}$ are $\propto (n^{ff})^{2}$. This means that the source function for the shocked region will be different for any elemental material at different locations within the jet volume. Hence we need to proceed by finding the quantities which are required to calculate the flux from any elemental material at an arbitrary location within the jet.


\begin{figure}
	\hskip -0.4cm
	\includegraphics[width=\columnwidth]{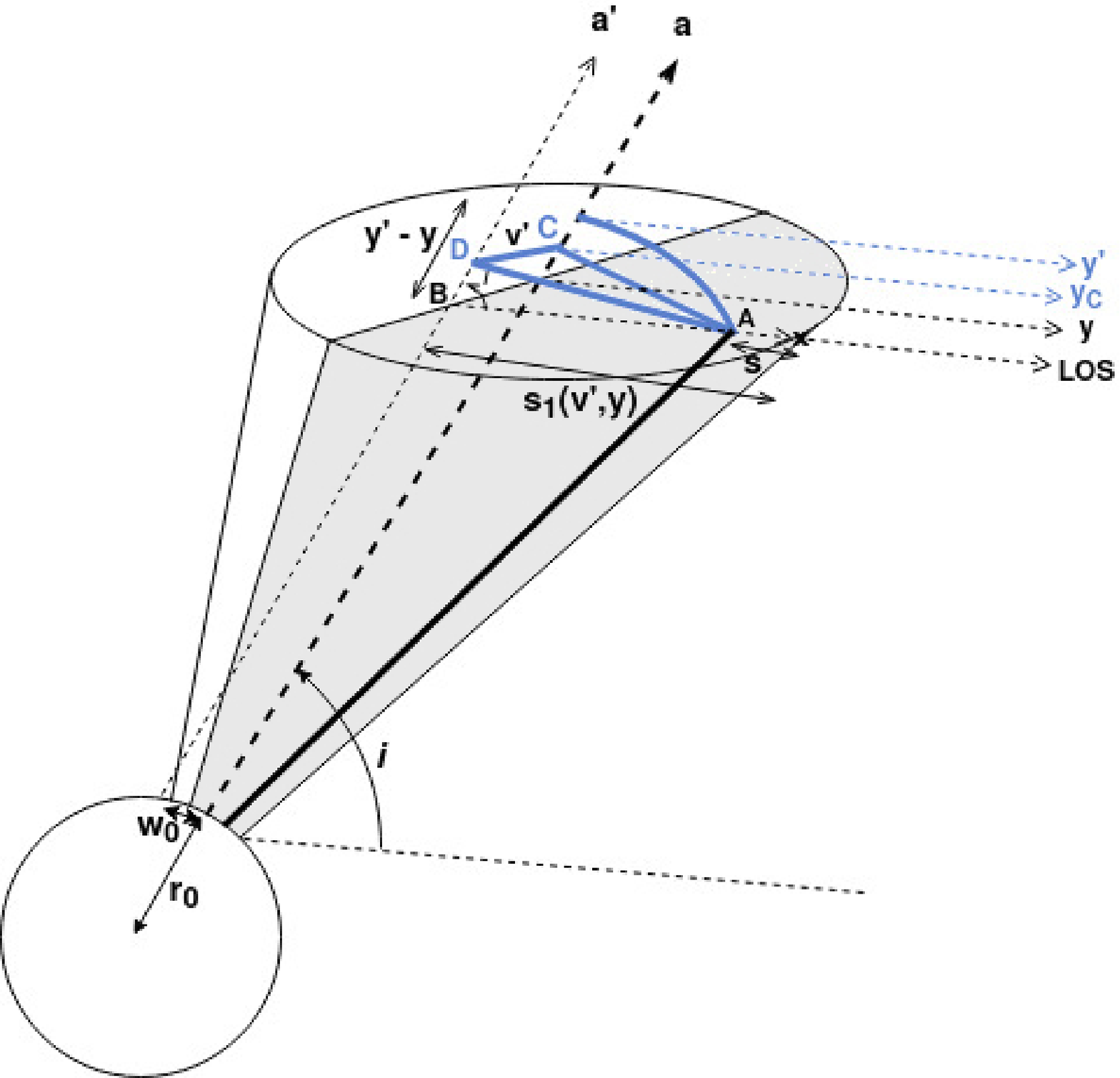}
    \caption{Side view of the jet, inclined at an angle $i$, slightly offset from the direction of the observer aligned at a length $y$ on the main axis. An arbitrary LOS located at $v'$ from the main axis along the sky plane is shown. The grey region represents the front half of the jet cross-section where the plane containing all the LOSs at $y$ cuts the jet. The blue lines (solid and dashed) are located above the LOS plane. y$_C$ is the projected length corresponding to the radial distance from $r_{a}=0$ along the jet axis upto C. $s$ and $s_1$($v'$,$y$) are the variable along the LOS and the full LOS distance for the front half of the jet at $v'$, respectively.}
    \label{fig:jet_geom}
\end{figure}

It is evident from Eqns.~\eqref{eq:shell_ff} and \eqref{eq:shell_nt} that for any element of material located at a distance $s$ along a LOS, the knowledge of $y'$, $w(y')$ and $w'(y')$ corresponding to its location are required to calculate its flux contribution. These elemental contributions can then be integrated first along each LOS, then the width and finally over the projected length to get the overall flux from the jet. Consider a jet as shown in Fig.~\ref{fig:jet_geom} with a constant opening angle ($\epsilon = 1$) whose main axis is $\vec{a}$ and the observer is located at a projected distance $y$ on the axis. Assume the plane formed by all the LOSs at $y$ which chops the jet at $y$, to be the LOS plane. The front half of this cross-section is represented as grey and all the regions above the LOS plane are represented as blue lines in the figure. Let us begin by assuming an arbitrary LOS in the LOS plane at a distance of $v'$ from the main axis such that $v'$ ranges from $0$ to $w(y)$ on either side of the axis. Note that $v'$ is the distance measured in the plane of the sky on either side of the axis, so the limits are slightly different for the top portion of the jet, where the curved edges of the jet dictate the projected width of the jet. Since the remaining calculations are the same, we do not consider the top portion for now. We know that the actual half-width corresponding to $y$ is $w(y)$. This is the half-width of the jet at $y$ when it is not inclined. The half-width of the jet at $v'$ is the half-chord at distance $v'$ from the axis given by,
\begin{equation}
w(v',y) = \sqrt{ w(y)^{2} - v'^{2}}
\end{equation} 
Using the relationship shown in Eqn.~\eqref{eq:s1}, the actual half width $w(v',y)$ can be used to calculate the projected widths on the front side and back side of the jet at $v'$ as,
\begin{equation}
s_{1}(v',y) = w(v',y)\, \frac{\cos{(\theta/2)}}{\sin{(i-\theta/2)}}
\end{equation}

\begin{equation}
s_{2}(v',y) = w(v',y)\, \frac{\cos{(\theta/2)}}{\sin{(i+\theta/2)}}
\end{equation}

Next, assume an elemental material located at A on the LOS at an arbitrary distance $s$ from the jet edge such that $s$ can vary from $0$ to $s_1(v',y)$. Here $s = s_1(v',y)$ corresponds to the point marked B in Fig.~\ref{fig:jet_geom}. In order to make the calculations simpler we project the axis $\vec{a}$, into the plane containing the points A, B and D, as axis $\vec{a'}$. AC and AD represent the distance of A from $\vec{a}$ and $\vec{a'}$, respectively, where C and D are separated by a distance of $v'$. The projected length of the jet axis upto C ($y_{C}$) can be calculated as,

\begin{equation} \label{tau}
{
\begin{split}
y_{C}- y & = (s_1(v',y)\,-\,s) \cos(i)\,\sin(i)\\
 \Rightarrow y_{C} & = y\,+\,(s_1(v',y)\,-\,s) \cos(i)\,\sin(i)
\end{split}
}
\end{equation}

The distance of A from the axis, $w'(s,v',y)$ ($\equiv$ AC) can be calculated from the geometry as,
\begin{equation} \label{w'}
w'(s,v',y) = \sqrt{((s_1(v',y)\,-\,s)\sin(i))^{2} + v'^{2}}
\end{equation}

From Eqns~\ref{tau} and~\ref{w'}, the projected length of the jet corresponding to A can be calculated as,
\begin{equation} 
y' = \sqrt{ y_{C}^{2} + {[w'(s,v',y)\sin(i)]}^{2}}
\end{equation}

Similar calculations can be done for the behind half of the jet. These quantities can then be substituted in Eqns.~\eqref{eq:shell_ff} and \eqref{eq:shell_nt} to find the number density of the electrons emitting the radiation. This helps in calculating the emission, absorption coefficients and optical depth which are required to obtain its flux contribution. Hence, the total radiation intensity for any LOS at $v'$ can be calculated as (similar to Eqn.~\eqref{eq:total_intensityI}),
\begin{equation}
\begin{split}
I_{\nu}(v',y) &= I_{\nu,R1}(v',y)\,e^{-(\tau^{ff}_{\nu,R2}(v',y)\,+\, \tau^{ff+syn}_{\nu,R3}(v',y))}\,\\
& +\, I_{\nu,R2}(v',y)\,e^{-\tau^{ff+syn}_{\nu,R3}(v',y)}\,+\, I_{\nu,R3}(v',y) 
\end{split}
\end{equation}

For a jet with varying opening angle ($\epsilon < 1$), the above calculations are similar except that for a given LOS, the opening angle corresponding to the front edge and the behind edge will be different as described in Appendix~\ref{ap:varyingth}, hence $s_1(v',y)$ and $s_2(v',y)$ are given by the Eqns.~\eqref{eq:s1_curved} and \eqref{eq:s2_curved}. 

This is extended to the top portion of the jet, from $y_{\text{LOS}_{\text{max}}}$ to $y_{max}$, with appropriate limits for $v'$ obtained from the geometrical calculations. Here, $s_1$ and $s_2$ are represented as $s_1^{top}$ and $s_2^{top}$, respectively. These are given by the following equations.

\begin{equation}
{
\begin{rcases}
  s_{1}^{top}(v',y) &= \frac{(\text{Y}_{\text{jet}} - y)}{\sin(i) \cos(i)} \\
  s_{2}^{top}(v',y) &= s_{2}(v',y) 
\end{rcases}
y_{\text{LOS}_{\text{max}}} \leq y < \text{Y}_{\text{jet}}}
\end{equation}

\begin{equation}
{
\begin{rcases}
  s_{1}^{top}(v',y) &= 0 \\
  s_{2}^{top}(v',y) &= s_2(v',y) - \frac{(y - \text{Y}_{\text{jet}})}{\sin(i) \cos(i)} 
\end{rcases}
\text{Y}_{\text{jet}} \leq y < y_{max}}
\end{equation}

We use these values to integrate the intensity along every LOS in the top portion.

\end{document}